%% file: Cut-off_rate_RIS_final.tex
\documentclass[twocolumn,journal]{IEEEtran}
\usepackage[T1]{fontenc}
\usepackage[latin9]{inputenc}
\usepackage{color}
\usepackage{verbatim}
\usepackage{prettyref}
\usepackage{amsmath}
\usepackage{amsthm}
\usepackage{amssymb}
\usepackage{graphicx}
\usepackage[unicode=true,
 bookmarks=true,bookmarksnumbered=true,bookmarksopen=true,bookmarksopenlevel=1,
 breaklinks=false,pdfborder={0 0 0},pdfborderstyle={},backref=false,colorlinks=false]
 {hyperref}
\hypersetup{pdftitle={Your Title},
 pdfauthor={Your Name},
 pdfpagelayout=OneColumn, pdfnewwindow=true, pdfstartview=XYZ, plainpages=false}

\makeatletter
\theoremstyle{plain}
\newtheorem{thm}{\protect\theoremname}
\theoremstyle{plain}
\newtheorem{lem}[thm]{\protect\lemmaname}

\usepackage[caption=false,font=footnotesize]{subfig}
\usepackage{acronym}

\AtBeginDocument{\input{acro.tex}}
\AtBeginDocument{\input{acro.tex}}
\usepackage{algorithm,algpseudocode}

\DeclareMathOperator{\minimize}{minimize}
\DeclareMathOperator{\st}{subject~to}
\DeclareMathOperator{\diag}{diag}
\DeclareMathOperator{\tr}{Tr}
\DeclareMathOperator{\vect}{vec}

\usepackage{pgfplots}
\usepackage{grffile}
\pgfplotsset{compat=newest}
\usetikzlibrary{plotmarks}
\usepgfplotslibrary{patchplots}
\usepackage{amsmath}

\definecolor{mycolor1}{rgb}{1.00000,0.00000,1.00000}%

\@ifundefined{showcaptionsetup}{}{%
 \PassOptionsToPackage{caption=false}{subfig}}
\usepackage{subfig}
\makeatother

\providecommand{\lemmaname}{Lemma}
\providecommand{\theoremname}{Theorem}

\begin{document}
\title{Optimization of RIS-aided MIMO Systems via the Cutoff Rate}
\author{\textcolor{black}{Nemanja~Stefan~Perovi\'c,~\IEEEmembership{\textcolor{black}{Member,~IEEE,}}
Le-Nam~Tran,~\IEEEmembership{\textcolor{black}{Senior Member,~IEEE,}}\\Marco
Di Renzo,~\IEEEmembership{\textcolor{black}{Fellow,~IEEE,}} and~Mark~F.~Flanagan,~\IEEEmembership{\textcolor{black}{Senior~Member,~IEEE}}}\thanks{\textcolor{black}{The work of N. S. Perovi\'c and M. F. Flanagan
was funded by the Irish Research Council under grant number IRCLA/2017/209.
The work of L. N. Tran was supported in part by a Grant from Science
Foundation Ireland under Grant number 17/CDA/4786. The work of M.
Di Renzo was supported in part by the European Commission through
the H2020 ARIADNE project under grant agreement number 871464 and
through the H2020 RISE-6G project under grant agreement number 101017011.}}\textcolor{black}{}\thanks{\textcolor{black}{N. S. Perovi\'c, L. N. Tran, and M. F. Flanagan
are with School of Electrical and Electronic Engineering, University
College Dublin, Belfield, Dublin 4,} D04 V1W8\textcolor{black}{, Ireland
(Email: nemanja.stefan.perovic@ucd.ie, nam.tran@ucd.ie and mark.flanagan@ieee.org).}}\textcolor{black}{}\thanks{\textcolor{black}{M. Di Renzo is with Universit\'e Paris-Saclay,
CNRS, CentraleSup\'elec, Laboratoire des Signaux et Syst\`emes,
3 Rue Joliot-Curie, 91192 Gif-sur-Yvette, France (E-mail: marco.di-renzo@universite-paris-saclay.fr).}}\textcolor{black}{\vspace{-0.1cm} }}
\maketitle
\begin{abstract}
The main difficulty concerning optimizing the \ac{MI} in \ac{RIS}-aided
communication systems with discrete signaling is the inability to
formulate this optimization problem in an analytically tractable manner.
Therefore, we propose to use the \ac{CR} as a more tractable metric
for optimizing the \ac{MI} and introduce two optimization methods
to maximize the CR,\textcolor{black}{{} assuming perfect knowledge of
the \ac{CSI}. The} first method is based on the \ac{PGM}, while
the second method is derived from the principles of \ac{SCA}\textcolor{black}{.}
Simulation results show that the proposed optimization methods significantly
enhance the CR and the corresponding \ac{MI}.\acresetall{}
\end{abstract}


\IEEEpeerreviewmaketitle{}

\section{Introduction}

\textcolor{black}{\bstctlcite{BSTcontrol}The recently developed
\acp{RIS} have the potential to shape and control the radio wave
propagation in wireless networks, which makes them a promising candidate
for future beyond-5G communication systems. \acp{RIS} consist of
a large number of small, low-cost, and nearly-passive elements each
of which can reflect the incident signal through an adjustable phase
shift, thereby modifying the wavefront of the scattered wave \cite{di2020smart}.
Changing the wavefront of the reflected signals enables us to shape
how the radio waves propagate through the channel, and thus improve
key system performance metrics such as the achievable rate. The main
body of research work in this area concentrates on the achievable
rate optimization for single-user \cite{wu2019intelligent} and multi-user
\cite{kammoun2020asymptotic} \ac{MISO} communication systems. Another
significant body of research work has focused on the achievable rate
optimization for \ac{MIMO} systems equipped with \acp{RIS} in single-user
\cite{perovic2019channel,perovic2020achievable} and multi-user \cite{pan2020multicell,pan2020intelligent}
communications. In the aforementioned papers, the transmitted symbols
are distributed according to a circularly-symmetric complex Gaussian
distribution, which is a capacity achieving distribution. However,
in practice the transmitted symbols are usually chosen from} a discrete
signal constellation and thus the present solutions cannot be used
to establish realistic bounds on the achievable data rate in practical
RIS-aided communication systems.

The main difficulty concerning the \emph{practical} achievable rate,
i.e., the analysis and optimization of the \ac{MI} in RIS-aided communication
systems with discrete signaling, is the inherent difficulty of formulating
this optimization problem in an analytically tractable manner. Hence,
\ac{MI} optimization and analysis were considered in only a few publications.
In \cite{karasik2020adaptive}, the authors considered the mutual
information optimization for an RIS-aided system, where the transmit
information is encoded into \ac{IQ} symbols transmitted from a single
transmit antenna and also into the (discrete) RIS phase shifts. 
However, the proposed optimization approach in \cite{karasik2020adaptive}
is not directly implementable to \ac{MIMO} systems that transmit
multiple \ac{IQ} symbols in parallel. The mutual information analysis
for a \ac{MISO} communication system, where the transmit information
is encoded into the \ac{IQ} symbol and a subset of active RIS elements,
is presented in \cite{lin2020reconfigurable}. The receiver device
is equipped with a single receive antenna in \cite{lin2020reconfigurable},
so the proposed optimization method is not applicable to multi-stream
\ac{MIMO} systems.

Motivated by this, we propose to use the channel \ac{CR} as a surrogate
metric for optimizing the \ac{MI} of RIS-aided multi-stream \ac{MIMO}
communication systems, \textcolor{black}{assuming perfect knowledge
of the \ac{CSI}.} The idea comes from the fact that the channel \ac{CR}
serves as a practical upper limit on the information rate for reliable
communications \cite{Massey1974}. In mathematical terms, the relation
between the channel \ac{CR} $R_{0}$ and the codeword error probability
$P_{e}$ can be formulated as $P_{e}\le e^{-n(R_{0}-R)}$, where $n$
is the codeword length and $R$ is the information rate in \ac{bpcu}.
In other words, for very long coded sequences (i.e., $n\rightarrow\infty$),
$P_{e}$ can be made arbitrarily small as long as $R<R_{0}$. Since
the channel capacity is the theoretical upper limit of the information
rate for reliable communications, the \ac{CR} is usually employed
as a practical lower bound on the channel capacity. More specifically,
it can be shown that a lower bound on the MI is determined by the
CR as \cite[Eq. (36)]{perovic2018optimization}
\[
\text{\ensuremath{\mathrm{MI}}}\ge N_{r}(1-\log_{2}e)+R_{0}^{h}
\]
where $R_{0}^{h}$ corresponds to the CR evaluated at half the noise
power. In practical terms, the use of the channel \ac{CR} facilitates
the otherwise intractable optimization of the channel capacity (i.e.,
\ac{MI}) and provides us with a tool to optimize modulation techniques
for communication systems. 

Against this background, the contributions of this paper are listed
as follows:
\begin{enumerate}
\item Instead of optimizing the MI directly, which results in an intractable
problem, we propose to use the \ac{CR} as a surrogate metric for
the MI optimization.  We show that the \ac{CR} can lead to a deterministic
optimization problem for which efficient numerical algorithms can
be derived.
\item To maximize the CR for the considered system, we formulate a joint
optimization problem of the precoding matrix and the RIS elements'
phase shifts. Since the problem of interest is nonconvex and our numerical
experiments indicates that it contains many local optima (see Section
\ref{sec:Simulation-Results}), we propose two local optimization
methods for solving this problem. The first method is a \ac{PGM}
which admits closed-form expressions in each iteration. The second
method is derived from the principles of \ac{SCA}. \textcolor{black}{To
achieve this, we prove that the non-convex unit-modulus constraints
on the RIS elements can be convexified without loss of optimality.
Hence,} one of the proposed methods can avoid being trapped in an
unsuitable local optimum, and thus will provide a near-optimal solution
for a given channel~realization.
\item We present simulation results which show that the proposed optimization
methods, by maximizing the \ac{CR}, can substantially increase the
\ac{MI}. \textcolor{black}{Moreover, the CR optimization produces
superior MI results compared to the achievable rate optimization for
Gaussian signaling.}
\end{enumerate}

\section{System Model}

\subsection{System Model}

\begin{figure}[t]
\centering{}\includegraphics[scale=0.65]{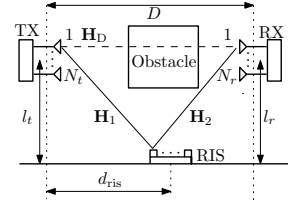}\caption{Aerial view of the considered communication system. \label{fig:Aerial-view}}
\end{figure}

An aerial view of the considered system is shown in \mbox{Fig. \ref{fig:Aerial-view}}.
It contains a transmitter equipped with $N_{t}$ antennas and a receiver
equipped with $N_{r}$ antennas, where we assume $N_{t}\ge N_{r}.$
The separations between the adjacent antennas in the transmit and
the receive antenna array are $s_{t}$ and $s_{r}$, respectively.
These antenna arrays are placed on parallel vertical walls, which
are at a distance $D$ from each other\textcolor{black}{}\footnote{\textcolor{black}{This system geometry described in the paper is adopted
for ease of exposition, and to provide the reader with a concrete
use case. However, the optimization approach proposed in this letter
is applicable to any system geometry.}}. To mitigate the blockage (i.e., attenuation) of the direct link,
an RIS, whose midpoint is at a distance $d_{\mathrm{ris}}$ from the
plane containing the transmit antenna array, is installed. The RIS
is placed on a vertical wall that is perpendicular to both the transmit
and the receive antenna array. We assume that the RIS, the transmit
antenna array and the receive antenna array are approximately at the
same height. The RIS consists of $N_{\mathrm{ris}}$ reflecting elements,
which are placed in a rectangular formation such that the separation
between the centers of adjacent RIS elements in both dimensions is
$s_{\mathrm{ris}}=\frac{\lambda}{2}$, where $\lambda$ denotes the
wavelength. The distance between the midpoint of the transmit (respectively,
receive) antenna array and the plane containing the RIS is $l_{t}$
(respectively, $l_{r}$).

The signal vector at the receive antenna array is given by
\begin{equation}
\mathbf{y}=\mathbf{HPx}_{i}+\mathbf{n},\label{eq:ss_equ-1}
\end{equation}
where $\mathbf{H}\in\mathbb{C}^{N_{r}\times N_{t}}$ is the channel
matrix, $\mathbf{P}\in\mathbb{C}^{N_{t}\times N_{r}}$ is the transmit
precoding matrix and $\mathbf{x}_{i}\in\mathbb{C}^{N_{r}\times1}$
is the transmit symbol vector. The elements of $\mathbf{x}_{i}$ are
chosen from a discrete symbol alphabet of size $M$ with unit average
symbol energy. The number of different transmit symbol vectors is
$N_{s}=M^{N_{r}}$. We assume that the precoding matrix~$\mathbf{P}$
preserves the average power of the transmit signal (i.e., \mbox{$\mathrm{Tr}(\mathbf{PP}^{H})=N_{r}$}).
The noise vector $\mathbf{n}\in\mathbb{C}^{N_{r}\times1}$ consists
of \ac{iid} elements that are distributed according to $\mathcal{CN}(0,\sigma^{2})$,
where $\sigma^{2}$ denotes the noise variance.

\textcolor{black}{Since an RIS is present in this system, the channel
matrix can be expressed as $\mathbf{H}=\sqrt{\beta_{\mathrm{DIR}}^{-1}}\mathbf{H}_{\mathrm{D}}+\sqrt{\beta_{\mathrm{INDIR}}^{-1}}\mathbf{H}_{2}\mathbf{F}(\boldsymbol{\theta})\mathbf{H}_{1}$,
where $\mathbf{H}_{\mathrm{D}}\in\mathbb{C}^{N_{r}\times N_{t}}$
represents the }\textcolor{black}{\emph{direct }}\textcolor{black}{link
between the transmitter and the receiver, $\mathbf{H}_{1}\in\mathbb{C}^{N_{\mathrm{ris}}\times N_{t}}$
represents the link between the transmitter and the RIS, and $\mathbf{H}_{2}\in\mathbb{C}^{N_{r}\times N_{\mathrm{ris}}}$
represents the link between the RIS and the receiver. The distance-dependent
path loss for the direct link is $\beta_{\mathrm{DIR}}^{-1}$ and
the \ac{FSPL} for the indirect link is $\beta_{\mathrm{INDIR}}^{-1}$.
Signal reflection from the RIS is modeled by $\mathbf{F}(\boldsymbol{\theta})=\mathrm{diag}(\boldsymbol{\theta})\in\mathbb{C}^{N_{\mathrm{ris}}\times N_{\mathrm{ris}}}$,
where $\boldsymbol{\theta}=[\theta_{1},\theta_{2},\ldots,\theta_{N_{\mathrm{ris}}}]^{T}\in\mathbb{C}^{N_{\mathrm{ris}}\times1}$.
Since we assume that the reflection is without any power loss, we
may write $\theta_{l}=e^{j\phi_{l}}$ (i.e., $\left|\theta_{l}\right|=1$)
for $l=1,2,\ldots,N_{\mathrm{ris}}$, where $\phi_{l}$ is the phase
shift induced by the $l$-th RIS element. In this letter, }\textit{\textcolor{black}{\emph{we
assume perfect knowledge of the \ac{CSI}. }}}

\section{Problem Formulation}

Since the considered system is a \ac{DCMC}, the \ac{MI} is given
by \cite[Eq. (7.2.1)]{gallager1968information}
\begin{equation}
\mathrm{MI}=\sum\nolimits _{i=1}^{N_{s}}\int_{\mathbf{y}}p(\mathbf{y},\mathbf{x}_{i})\log_{2}\Bigl[\frac{p(\mathbf{y}|\mathbf{x}_{i})}{\sum_{j=1}^{N_{s}}p(\mathbf{y},\mathbf{x}_{j})}\Bigr]d\mathbf{y},
\end{equation}
where the conditional probability density function is 
\begin{equation}
p(\mathbf{y}|\mathbf{x}_{i})=\Bigl(1/\left(\pi\sigma^{2}\right)^{N_{r}}\Bigr)\text{\ensuremath{\exp}}\Bigl(-\frac{\left\Vert \mathbf{y}-\mathbf{HP}\mathbf{x}_{i}\right\Vert ^{2}}{\sigma^{2}}\Bigr).\label{eq:pdf}
\end{equation}
Assuming that all the transmit symbol vectors are equally probable
$p(\mathbf{x}_{i})=1/N_{s}$ $(i=1,\dots,N_{s})$, we obtain
\begin{equation}
\mathrm{MI}=\log_{2}N_{s}-\frac{1}{N_{s}}\sum\nolimits _{i=1}^{N_{s}}\mathbb{E}_{\mathbf{n}}\Bigl\{\log_{2}\sum\nolimits _{j=1}^{N_{s}}\exp(\psi_{i,j})\Bigr\},\label{eq:cap}
\end{equation}
where $\psi_{i,j}=(-\left\Vert \mathbf{HP}(\mathbf{x}_{i}-\mathbf{x}_{j})+\mathbf{n}\right\Vert ^{2}+\left\Vert \mathbf{n}\right\Vert ^{2})/\sigma^{2}$.

We remark that the expression for the MI in \eqref{eq:cap} leads
to an intractable stochastic optimization problem, if one wishes to
maximize the MI directly. The fact that the MI expression in \eqref{eq:cap}
is neither convex nor concave with respect to $\mathbf{P}$ and $\boldsymbol{\theta}$
also adds to the difficulty. Also, the feasible set for $\theta_{l}$,
which satisfies $\left|\theta_{l}\right|=1$, is nonconvex. To overcome
these issues, we propose the use of the \ac{CR} as an auxiliary metric
whose optimization is well-aligned with that of the MI.

\subsection{Derivation of the \ac{CR}}

The \ac{CR} of the considered system for equiprobable symbol vectors
$\mathbf{x}_{i}$ is given by \cite{Massey1974}
\begin{equation}
R_{0}=-\text{\ensuremath{\log}}_{2}\Bigl[\frac{1}{N_{s}^{2}}\int_{\mathbf{y}}\sum\nolimits _{i,j=1}^{N_{s}}\sqrt{p(\mathbf{y}|\mathbf{x}_{i})p(\mathbf{y}|\mathbf{x}_{j})}d\mathbf{y}\Bigr].\label{eq:cut_off}
\end{equation}
By inserting \prettyref{eq:pdf} into \prettyref{eq:cut_off} and
after some algebraic manipulations, the \ac{CR} of the considered
system can be written as
\begin{equation}
R_{\text{0}}=-\text{\ensuremath{\log}}_{2}\biggl[\frac{1}{N_{s}^{2}}\sum\nolimits _{i,j=1}^{N_{s}}\exp\Bigl(-\frac{\Phi_{i,j}(\boldsymbol{\theta},\mathbf{P})}{4\sigma^{2}}\Bigr)\biggr],\label{eq:cut-off-fin}
\end{equation}
where $\Phi_{i,j}(\boldsymbol{\theta},\mathbf{P})=\left\Vert \mathbf{HP}\mathbf{e}_{i,j}\right\Vert ^{2}$
and $\mathbf{e}_{i,j}=\mathbf{x}_{i}-\mathbf{x}_{j}$.

\subsection{Optimization Problem}

Upon close inspection of \eqref{eq:cut-off-fin}, it is easy to see
that the \ac{CR} optimization problem can be formulated as \begin{subequations}\label{subsec:CR_opt_prob}
\begin{align}
\underset{\boldsymbol{\theta},\mathbf{P}}{\minimize}\  & f(\boldsymbol{\theta},\mathbf{P})=\sum\nolimits _{i,j=1}^{N_{s}}\exp\Bigl(-\frac{\Phi_{i,j}(\boldsymbol{\theta},\mathbf{P})}{4\sigma^{2}}\Bigr)\\
\st\  & \mathrm{Tr}(\mathbf{PP}^{H})=N_{r}\label{eq:powerconstraint}\\
 & \bigr|\theta_{l}\bigr|=1,l=1,2,\ldots,N_{\mathrm{ris}}.\label{eq:phaseconstraint}
\end{align}
\end{subequations}Since the objective of \eqref{subsec:CR_opt_prob}
is nonconvex%
, finding a globally optimal solution is very difficult. Hence, in
the next section, we present two local optimization algorithms to
solve \eqref{subsec:CR_opt_prob}.\vspace{-0.5em}

\section{Proposed Optimization Methods}

\subsection{Projected Gradient Method (PGM)}

The first proposed method is based on the \ac{PGM}%
{} \cite{boyd2004convex}, which consists of the following iterations:\begin{subequations}
\begin{align}
\boldsymbol{\theta}_{n+1} & =P_{\Theta}(\boldsymbol{\theta}_{n}-\mu_{1}\nabla_{\boldsymbol{\theta}}f(\boldsymbol{\theta}_{n},\mathbf{P}_{n})),\label{eq:Calc_thata}\\
\mathbf{P}_{n+1} & =P_{\mathcal{P}}(\mathbf{P}_{n}-\mu_{2}\nabla_{\mathbf{P}}f(\boldsymbol{\theta}_{n+1},\mathbf{P}_{n})),\label{eq:Calc_P}
\end{align}
\end{subequations}where $\nabla_{\boldsymbol{\theta}}f(\boldsymbol{\theta},\mathbf{P})$
and $\nabla_{\mathbf{P}}f(\boldsymbol{\theta},\mathbf{P})$ are the
gradients of $f(\boldsymbol{\theta},\mathbf{P})$ with respect to
$\boldsymbol{\theta}^{\ast}$ and $\mathbf{P}^{\ast}$, respectively\footnote{Here $x^{*}$ denotes the complex conjugate of $x$.},
and $\mu_{1}$ and $\mu_{2}$ are the corresponding step sizes. Also,
$P_{\Theta}(\cdot)$ and $P_{\mathcal{P}}(\cdot)$ denote the projection
onto $\Theta$ and $\mathcal{P}$, respectively, which are detailed
in the sequel. Note that we use the complex-valued gradients~\cite{Are2011},
the explicit forms of which are provided in the next lemma.
\begin{lem}
The gradients of $f(\boldsymbol{\theta},\mathbf{P})$ with respect
to $\boldsymbol{\theta}^{\ast}$ and $\mathbf{P}^{\ast}$ are given
by{\small 
\begin{subequations} 	
\begin{IEEEeqnarray}{rcl} 		
\nabla_{\boldsymbol{\theta}}f(\boldsymbol{\theta},\mathbf{P}) &=&-\frac{1}{4\sigma^{2}}\sum_{i,j=1}^{N_{s}}e^{-\frac{\Phi_{i,j}(\boldsymbol{\theta},\mathbf{P})}{4\sigma^{2}}} \vect_{d}(\mathbf{H}_{2}^{H}\mathbf{H}\mathbf{P}_{i,j}\mathbf{P}_{i,j}^{H}\bar{\mathbf{H}}_{1}^{H})\label{eq:grad_theta}\IEEEeqnarraynumspace\vspace{-8pt}\\  		\nabla_{\mathbf{P}}f(\boldsymbol{\theta},\mathbf{P})&=&-\frac{1}{4\sigma^{2}}\mathbf{H}^{H}\mathbf{HP}\sum_{i,j=1}^{N_{s}}e^{-\frac{\Phi_{i,j}(\boldsymbol{\theta},\mathbf{P})}{4\sigma^{2}}}\mathbf{e}_{i,j}\mathbf{e}_{i,j}^{H},\label{eq:grad_P} 	
\end{IEEEeqnarray} 
\end{subequations} 
}where $\mathbf{P}_{i,j}=\mathbf{Pe}_{i,j}$, $\bar{\mathbf{H}}_{1}=\sqrt{\beta_{\mathrm{INDIR}}^{-1}}\mathbf{H}_{1}$
and $\vect_{d}(\mathbf{A})$ denotes the vector comprised of the diagonal
elements of $\mathbf{A}$.
\end{lem}
\begin{IEEEproof}
See Appendix A.
\end{IEEEproof}

\subsubsection{Projection Operations}

Next we show that the projection operations in \eqref{eq:Calc_thata}
and \eqref{eq:Calc_P} can be calculated in closed form. Note that
the constraint $\bigl|\theta_{l}\bigr|=1$ states that $\theta_{l}$
lies on the unit circle in the complex plane. Thus, for a given point
$\boldsymbol{\theta}\in\mathbb{C}^{N_{\mathrm{ris}}\times1}$, $P_{\Theta}(\mathbf{\boldsymbol{\theta}})$
is given by%
\begin{equation}
\bar{\theta}_{l}=\begin{cases}
\frac{\theta_{l}}{|\theta_{l}|} & \theta_{l}\neq0\\
e^{j\phi},\phi\in[0,2\pi] & \theta_{l}=0
\end{cases},\;l=1,\dots,N_{\mathrm{ris}}.\label{eq:projectthetha}
\end{equation}
In particular, $\bar{\theta}_{l}$ can be any point on the unit circle
if $\theta_{l}=0$, and thus $P_{\Theta}(\mathbf{\boldsymbol{\theta}})$
is not unique.

Similarly, the constraint $\mathrm{Tr}(\mathbf{PP}^{H})=N_{r}$ implies
that the projection of the precoding matrix $P_{\mathcal{P}}(\mathbf{P})$
is given by
\begin{equation}
\bar{\mathbf{P}}=\mathbf{P}\sqrt{N_{r}/\mathrm{Tr}(\mathbf{PP}^{H})}.
\end{equation}

\subsubsection{Backtracking Line Search}

Appropriate choices of the step sizes in \eqref{eq:Calc_thata} and
\eqref{eq:Calc_P} are instrumental for ensuring the convergence of
the PGM. Ideally, each step size should be inversely proportional
to the Lipschitz constant of the corresponding gradient. However,
the optimal Lipschitz constants are difficult to find for our considered
problem. Therefore, we utilize the Armijo-Goldstein backtracking line
search to determine the step sizes $\mu_{1}$ and $\mu_{2}$ at each
iteration.

Let $L_{0}>0$, $\delta>0$ be a small constant, and $\rho\in(0,1)$.
The step size $\mu_{1}$ in \eqref{eq:Calc_thata} is found as $L_{o}\rho^{k_{n}}$,
where $k_{n}$ is the smallest nonnegative integer such that
\begin{equation}
f(\boldsymbol{\theta}_{n+1},\mathbf{P}_{n})\leq f(\boldsymbol{\theta}_{n},\mathbf{P}_{n})-\delta||\boldsymbol{\theta}_{n+1}-\boldsymbol{\theta}_{n}||^{2}.
\end{equation}
The step size $\mu_{2}$ is found similarly. The proposed line search
procedure ensures that the objective sequence strictly decreases after
each iteration. Thus, the PGM is guaranteed to converge to a stationary
point of \eqref{subsec:CR_opt_prob}, which is, however, not necessarily
a globally optimal solution.

\subsection{Successive Convex Approximation (SCA)}

Since \eqref{subsec:CR_opt_prob} is a nonconvex problem, the proposed
PGM can become trapped in an unsuitable local optimum. Thus, it is
desirable to check the obtained solution with another local optimization
method that is developed by applying a different optimization paradigm.
Our expectation is that the two proposed local optimization methods
can complement each other and one of them can escape unsuitable local
optima. In particular, the second proposed method is derived from
the \ac{SCA}, which has been shown to be a powerful tool for a range
of nonconvex optimization problems. %

First, we relax the equality constraints in \eqref{subsec:CR_opt_prob},
leading to the following program: \begin{subequations}\label{eq:rewrite}
\begin{align}
\minimize & \ f(\boldsymbol{\theta},\mathbf{P})=\sum\nolimits _{i,j=1}^{N_{s}}\exp\left(-\frac{\Phi_{i,j}(\boldsymbol{\theta},\mathbf{P})}{4\sigma^{2}}\right)\\
\st & \ \mathrm{Tr}(\mathbf{PP}^{H})\leq N_{r}\label{eq:relaxpower}\\
 & \ \left|\theta_{l}\right|\leq1,l=1,2,\ldots,N_{\textrm{ris}}.\label{eq:relaxphase}
\end{align}
\end{subequations}It is easy to see that $\Phi_{i,j}(\boldsymbol{\theta},\alpha\mathbf{P})>\Phi_{i,j}(\boldsymbol{\theta},\mathbf{P})$
for any $\alpha>1$. Thus, the constraint in \eqref{eq:relaxpower}
must hold with equality at the optimum. Otherwise, we can always scale
up $\mathbf{P}$ and achieve a strictly smaller objective. Similarly,
it can be shown that \eqref{eq:relaxphase} holds with equality at
the optimum (see Appendix \ref{sec:Relax_app}). These two observations
imply that the set of optimal solutions of \eqref{eq:rewrite} is
the same as that of \eqref{subsec:CR_opt_prob}. Also, the aforementioned
relaxation ensures the convexity of the feasible sets for $\boldsymbol{\theta}$
and $\mathbf{P}$, while the objective function $f(\boldsymbol{\theta},\mathbf{P})$
remains nonconvex.

Let us suppose that we have already obtained a feasible point $(\mathbf{P}_{n},\boldsymbol{\theta}_{n})$.
To find $\mathbf{P}_{n+1}$ based on the SCA method, we need to find
a convex upper bound on $f(\boldsymbol{\theta}_{n},\mathbf{P})$.
Since $\Phi_{i,j}(\boldsymbol{\theta}_{n},\mathbf{P})$ is convex
with respect to $\mathbf{P}$, we have 
\begin{align}
\Phi_{i,j}(\boldsymbol{\theta}_{n},\mathbf{P}) & \geq\Phi_{i,j}(\boldsymbol{\theta}_{n},\mathbf{P}_{n})+\bigl\langle\nabla_{\mathbf{P}}\Phi_{i,j}(\boldsymbol{\theta}_{n},\mathbf{P}_{n}),\mathbf{P}-\mathbf{P}_{n}\bigr\rangle\nonumber \\
\stackrel{\triangle}{=} & \tilde{\Phi}_{i,j}(\boldsymbol{\theta}_{n},\mathbf{P}_{n};\mathbf{P})\label{eq:P_approx}
\end{align}
where $\bigl\langle\mathbf{X},\mathbf{Y}\bigr\rangle=2\Re(\tr(\mathbf{X}^{H}\mathbf{Y}))$.
In fact, $\tilde{\Phi}_{i,j}(\boldsymbol{\theta}_{n},\mathbf{P}_{n};\mathbf{P})$
is an affine approximation of $\Phi_{i,j}(\boldsymbol{\theta}_{n},\mathbf{P})$
around $\mathbf{P}_{n}$. Next, $\mathbf{P}_{n+1}$ is a solution
to the following convex problem:{\small
\begin{equation}
\underset{\mathbf{P}}{\min}\Bigl\{\sum\nolimits _{i,j=1}^{N_{s}}\exp\Bigl(-\frac{\tilde{\Phi}_{i,j}(\boldsymbol{\theta}_{n},\mathbf{P}_{n};\mathbf{P})}{4\sigma^{2}}\Bigr)\:\Bigr|\:\mathrm{Tr}(\mathbf{PP}^{H})\leq N_{r}\Bigr\}.\label{eq:subprobP}
\end{equation}

}

To find $\boldsymbol{\theta}_{n+1}$, we again need to derive a convex
upper bound on $f(\boldsymbol{\theta},\mathbf{P}_{n+1})$. It is easy
to see that $\Phi_{i,j}(\boldsymbol{\theta},\mathbf{P}_{n+1})$ is
convex with respect to $\boldsymbol{\theta}$ and thus we have 
\begin{gather}
\Phi_{i,j}(\boldsymbol{\theta},\mathbf{P}_{n+1})\geq\Phi_{i,j}(\boldsymbol{\theta}_{n},\mathbf{P}_{n+1})+\nonumber \\
\bigl\langle\nabla_{\boldsymbol{\theta}}\Phi_{i,j}(\boldsymbol{\theta}_{n},\mathbf{P}_{n+1}),\boldsymbol{\theta}-\boldsymbol{\theta}_{n}\bigr\rangle\stackrel{\triangle}{=}\bar{\Phi}_{i,j}(\boldsymbol{\theta}_{n},\mathbf{P}_{n+1};\boldsymbol{\theta}).\label{eq:theta_approx}
\end{gather}
Next, $\boldsymbol{\theta}_{n+1}$ is a solution of the following
convex problem:\begin{subequations}\label{eq:subprobtheta}
\begin{align}
\underset{\boldsymbol{\theta}}{\minimize} & \ \sum\nolimits _{i,j=1}^{N_{s}}\exp\Bigl(-\frac{\bar{\Phi}_{i,j}(\boldsymbol{\theta}_{n},\mathbf{P}_{n+1};\boldsymbol{\theta})}{4\sigma^{2}}\Bigr)\\
\st & \ \left|\theta_{l}\right|\leq1,l=1,2,\ldots,N_{\textrm{ris}}.
\end{align}
\end{subequations}Both convex optimization problems in \eqref{eq:subprobP}
and \eqref{eq:subprobtheta} can be solved by off-the-shelf convex
solvers. By minimizing an upper bound in each iteration, the SCA-based
method generates a decreasing objective sequence. It can shown that
the SCA-based method converges to a stationary point of \eqref{eq:rewrite}.

\subsection{\textcolor{black}{Complexity Comparison}}

\textcolor{black}{We next provide a brief analysis of the complexity
of the proposed methods based on the required number of complex multiplications.
The per-iteration complexity of the PGM method is $\mathcal{O}(N_{s}^{2}N_{\mathrm{ris}}(N_{t}+N_{r}))$.
To solve subproblems \eqref{eq:subprobP} and \eqref{eq:subprobtheta},
we first need to compute \eqref{eq:P_approx} and \eqref{eq:theta_approx},
and this requires approximately the same per-iteration complexity
as the PGM. By treating \eqref{eq:subprobP} and \eqref{eq:subprobtheta}
as generic convex problems, the worst-case complexity for solving
\eqref{eq:subprobP} and \eqref{eq:subprobtheta} is $\mathcal{O}(N_{t}^{4}N_{r}^{4})$
and $\mathcal{O}(N_{\textrm{ris}}^{4})$, respectively.}

\textcolor{black}{The SCA-based method generally requires higher computational
complexity to return a solution than the PGM method. The reason is
that convex solvers commonly use an interior point method to solve
problems \eqref{eq:subprobP} and \eqref{eq:subprobtheta} which involve
exponential cones. The memory requirement and the computational complexity
increase very quickly with the problem size. Consequently, the SCA-based
method is not suitable for large-scale scenarios, i.e., when $N_{s}$
or $N_{\textrm{ris}}$ is large. For such cases, the PGM is the only
viable option. For problems of moderate size, the SCA-based method
is a good choice since it is a descent method without a line search.
On the other hand, the PGM method requires a line search to ensure
a good convergence rate.}%

\section{Simulation Results\label{sec:Simulation-Results}}

In this section, we evaluate the \ac{CR} and the \ac{MI} of the
proposed optimization algorithms with the aid of Monte Carlo simulations.
More precisely, we utilize $\boldsymbol{\theta}$ and $\mathbf{P}$
obtained by optimizing the CR to calculate the MI according to the
expression in \eqref{eq:cap}, where the expectation is~evaluated
by generating random noise vectors. Also, we compare the~MI~optimization
results in the case of discrete and Gaussian signaling.

\textcolor{black}{In the following simulations, all of the channel
matrices are modeled according to the Rician fading channel model
with Rician factor $K$, as specified in \cite{perovic2020achievable}.
Also, we assume no spatial correlation}\footnote{\textcolor{black}{Spatial correlation models for RIS channel matrices
have recently become available in the literature (e.g., in \cite{bjornson2020rayleigh}).}}\textcolor{black}{{} exists among the elements of matrices $\mathbf{H}_{1}$
and $\mathbf{H}_{2}$. The distance-dependent path loss for th}e direct
link is $\beta_{\mathrm{DIR}}=(4\pi/\lambda)^{2}d_{0}^{\alpha_{\mathrm{DIR}}}$,
where $d_{0}$ is the distance between the transmit array midpoint
and the receive array midpoint, and the path loss exponent of the
direct link is denoted by $\alpha_{\mathrm{DIR}}$. The far-field
\ac{FSPL} for the indirect link $\beta_{\mathrm{INDIR}}^{-1}$ can
be computed according to \cite{danufane2020path} as $\beta_{\mathrm{INDIR}}^{-1}=\lambda^{4}(l_{t}/d_{1}+l_{r}/d_{2})^{2}/(256\pi^{2}d_{1}^{2}d_{2}^{2}),$
where $d_{1}$ is the distance between the transmit antenna array
midpoint and the RIS center, and $d_{2}$ is the distance between
the RIS center and the receive antenna array midpoint.\textcolor{black}{}
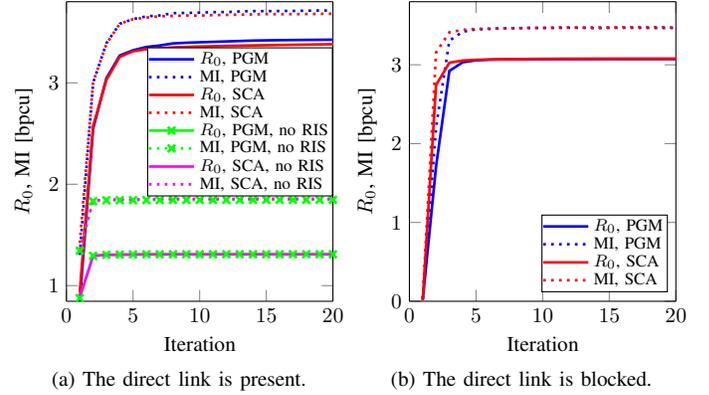
\begin{figure}[t]
\textcolor{black}{}\subfloat[The direct link is present.]{\textcolor{black}{\input{figures_scaled/plot_8_2_dir_sc_V1.tex}}}\textcolor{black}{}\subfloat[The direct link is blocked.]{\textcolor{black}{\input{figures_scaled/plot_8_2_nondir_sc.tex}}}

\textcolor{black}{\caption{The \ac{CR} ($R_{0}$) and the \ac{MI} of the proposed PGM and SCA
methods.\label{fig:CR_and_MU}}
}
\end{figure}

\textcolor{black}{In the following simulation setup, the parameters
are $f=2\,\mathrm{GHz}$ (i.e., $\lambda=15\,\mathrm{cm}$), $s_{t}=s_{r}=s_{\mathrm{ris}}=\lambda/2=7.5\,\mathrm{cm}$,
$D=500\,\mathrm{m}$, $l_{t}=l_{r}=20\,\mathrm{m},$ $d_{\mathrm{ris}}=30\,\mathrm{m}$,
$N_{t}=8$, $N_{r}=2$, $\alpha_{\mathrm{DIR}}=3$, $K=1$, $M=4$,
and $\sigma^{2}=-110\thinspace\mathrm{dB}$. The RIS consists of $N_{\mathrm{ris}}=225$
elements placed in a \mbox{$15\times15$} square formation. The
line search procedure for the PGM utilizes the parameters $L_{0}=10^{3}$,
$\delta=10^{-3}$ and $\rho=1/2$. For the SCA-based method, we use
the CVX tool with MOSEK as the internal software package to solve
\eqref{eq:subprobP} and~\eqref{eq:subprobtheta}. The initial values
of $\boldsymbol{\theta}$ and $\mathbf{P}$ are randomly chosen for
both methods. All results are averaged over 30 independent channel
realizations.}

\textcolor{black}{In Fig. \ref{fig:CR_and_MU}, we show the \ac{CR}
and the \ac{MI} results of the proposed PGM and SCA for scenarios
in which the direct link is present and in which the direct link is
blocked. As benchmark schemes, we consider the presented system without
the RIS, whose precoding matrix is optimized by using the proposed
methods. The two proposed optimization methods achieve approximately
the same results and need only a few iterations to reach the optimum.
In general, the MI shows the same behavioral trend as the CR, but
the MI is always larger than the CR. For both metrics, the proposed
algorithms achieve a similar gain through the parameter optimization,
which is around 2~bpcu and 3~bpcu if the direct link is present
and blocked, respectively. These results justify our initial claim
that the \ac{CR} presents a suitable metric to optimize~the~\ac{MI}.
In addition, removing the RIS from the considered system causes a
significant reduction of the CR and the MI.}

\begin{figure}[t]
\centering{}\subfloat[The direct link is present.]{\centering{}\input{figures_scaled/dis_vs_cont_dir_sc.tex}}\subfloat[The direct link is blocked.]{\centering{}\input{figures_scaled/dis_vs_cont_nondir_sc.tex}}\caption{MI optimization results for discrete and Gaussian signaling.\label{fig:MI-signaling}}
\end{figure}
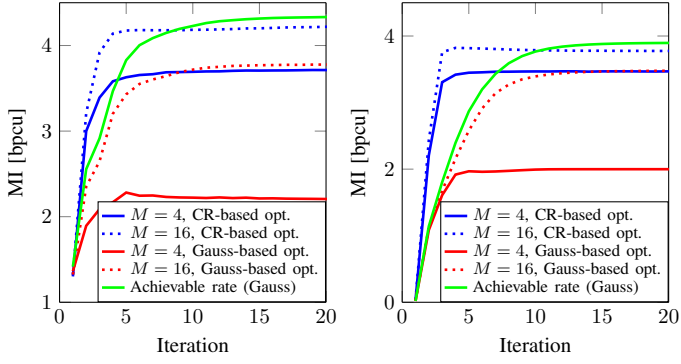

\textcolor{black}{In Fig. \ref{fig:MI-signaling}, we compare the
\ac{MI} when discrete and Gaussian signaling are considered. Specifically,
we plot the MI where the precoding matrix and the RIS coefficients
are obtained by implementing the CR optimization (referred to as CR-based
optimization), and the MI where the precoding matrix and the RIS coefficients
are obtained based on the achievable rate optimization for Gaussian
signaling (referred to as Gaussian signaling based optimization) \cite{perovic2020achievable}.
Moreover, we present the achievable rate for Gaussian signaling \cite{perovic2020achievable}.
As can be seen, the CR-based optimization yields a higher MI than
the Gaussian signaling based optimization, particularly for a small
size of symbol alphabet. As expected, the achievable rate for Gaussian
signaling is always higher than the MI. However, for $M=16$ the gap
between the MI obtained by using the discrete optimization and the
achievable rate is less than 0.2~bpcu. }

\section{Conclusion}

In this letter, we have investigated the use of the CR as a simple
and meaningful metric for optimizing the MI in RIS-aided MIMO communication
systems. Since the maximization of the CR is a nonconvex optimization
problem, we proposed two local optimization methods. The first method
is based on the PGM which uses closed-form expressions in each iteration.
The second method is derived from the principles of SCA. Simulation
results show that both optimization methods produce very similar performance
results and that the CR is indeed a suitable substitute metric for
optimizing the \ac{MI} in RIS-aided MIMO communication systems.\textcolor{black}{{}
This work can be extended to the case where the transmitted data is
simultaneously encoded into the \ac{IQ} symbols and the RIS phase
shifts, as well as to the case where }\textcolor{black}{\emph{statistical}}\textcolor{black}{{}
\ac{CSI} is used for the CR optimization. }

\appendices{}

\section{Complex-valued Gradient of $f(\boldsymbol{\theta},\mathbf{Q})$}

The complex gradient of $f(\boldsymbol{\theta},\mathbf{Q})$ with
respect to $\boldsymbol{\theta}^{\ast}$ is{\small
\begin{equation}
\!\!\!\!\!\!\nabla_{\boldsymbol{\theta}}f(\boldsymbol{\theta},\mathbf{P})=-\frac{1}{4\sigma^{2}}\sum_{i,j=1}^{N_{s}}\exp\Bigl[-\frac{\Phi_{i,j}(\boldsymbol{\theta},\mathbf{P})}{4\sigma^{2}}\Bigr]\nabla_{\boldsymbol{\theta}}\Phi_{i,j}(\boldsymbol{\theta},\mathbf{P}).\label{eq:grad_phi_1}
\end{equation}
}Also, the complex differential of $\Phi_{i,j}$ with respect to
$\mathbf{F}(\boldsymbol{\theta})=\diag(\boldsymbol{\theta})$ and
$\mathbf{F}^{*}(\boldsymbol{\theta})$ can be expressed as
\begin{align}
\!d\Phi_{i,j}(\boldsymbol{\theta},\mathbf{P}) & =d(\left\Vert \mathbf{HPe}_{i,j}\right\Vert ^{2})=\mathbf{e}_{i,j}^{H}\mathbf{P}^{H}d(\mathbf{H}^{H}\mathbf{H})\mathbf{Pe}_{i,j}.
\end{align}
With the aid of a few algebraic steps, we obtain
\begin{gather}
d\Phi_{i,j}(\boldsymbol{\theta},\mathbf{P})=\vect^{T}\left(\mathbf{H}_{2}^{H}\mathbf{H}\mathbf{Pe}_{i,j}\mathbf{e}_{i,j}^{H}\mathbf{P}^{H}\bar{\mathbf{H}}_{1}^{H}\right)\vect(d(\mathbf{F}^{*}))+\nonumber \\
\vect^{T}\left(\left(\bar{\mathbf{H}}_{1}^{H}\mathbf{Pe}_{i,j}\mathbf{e}_{i,j}^{H}\mathbf{P}^{H}\mathbf{H}^{H}\mathbf{H}_{2}\right)^{T}\right)\vect(d(\mathbf{F}))
\end{gather}
where we used the identity $\tr(\mathbf{A}^{T}\mathbf{B})=\vect^{T}(\mathbf{A})\vect(\mathbf{B})$,
where $\vect(\mathbf{A})$ denotes the vector obtained by vertical
stacking of the columns of $\mathbf{A}$. Let $\mathbf{L}_{d}$ be
the matrix used to place the diagonal elements of a square matrix
$\mathbf{A}$ on $\vect(\mathbf{A})$, i.e., $\vect(\mathbf{A})=\mathbf{L}_{d}\vect_{d}(\mathbf{A})$
\cite[Def. 2.12]{Are2011}. Then, we have
\begin{gather}
d\Phi_{i,j}(\boldsymbol{\theta},\mathbf{P})=\vect^{T}\left(\mathbf{H}_{2}^{H}\mathbf{H}\mathbf{Pe}_{i,j}\mathbf{e}_{i,j}^{H}\mathbf{P}^{H}\bar{\mathbf{H}}_{1}^{H}\right)\mathbf{L}_{d}\vect(d\boldsymbol{\theta}^{*})+\nonumber \\
\vect^{T}\left(\left(\bar{\mathbf{H}}_{1}^{H}\mathbf{Pe}_{i,j}\mathbf{e}_{i,j}^{H}\mathbf{P}^{H}\mathbf{H}^{H}\mathbf{H}_{2}\right)^{T}\right)\mathbf{L}_{d}\vect(d\boldsymbol{\theta}).
\end{gather}
Using \cite[Table 3.2]{Are2011} and \cite[Eqn. (2.140)]{Are2011},
we obtain
\begin{gather}
\nabla_{\boldsymbol{\theta}}\Phi_{i,j}(\boldsymbol{\theta},\mathbf{P})=\mathbf{L}_{d}^{T}\vect\left(\mathbf{H}_{2}^{H}\mathbf{H}\mathbf{Pe}_{i,j}\mathbf{e}_{i,j}^{H}\mathbf{P}^{H}\bar{\mathbf{H}}_{1}^{H}\right)\nonumber \\
=\vect_{d}\left(\mathbf{H}_{2}^{H}\mathbf{H}\mathbf{Pe}_{i,j}\mathbf{e}_{i,j}^{H}\mathbf{P}^{H}\bar{\mathbf{H}}_{1}^{H}\right).\label{eq:dif_phi_fin}
\end{gather}
Substituting \eqref{eq:dif_phi_fin} into \eqref{eq:grad_phi_1},
we obtain \eqref{eq:grad_theta}.

In a similar way, the gradient of $f(\boldsymbol{\theta},\mathbf{P})$
with respect to $\mathbf{P}^{\ast}$ is equal to{\small
\begin{equation}
\!\!\!\!\!\!\nabla_{\mathbf{P}}f(\boldsymbol{\theta},\mathbf{P})=-\frac{1}{4\sigma^{2}}\sum_{i,j=1}^{N_{s}}\exp\left(-\frac{\Phi_{i,j}(\boldsymbol{\theta},\mathbf{P})}{4\sigma^{2}}\right)\nabla_{\mathbf{P}}\Phi_{i,j}(\boldsymbol{\theta},\mathbf{P}).\label{eq:P_grad_1}
\end{equation}
}The complex differential of $\Phi_{i,j}$ with respect to $\mathbf{P}$
and $\mathbf{P}^{\ast}$is
\begin{gather}
d\Phi_{i,j}(\boldsymbol{\theta},\mathbf{P})=d(\mathbf{e}_{i,j}^{H}\mathbf{P}^{H}\mathbf{H}^{H}\mathbf{HPe}_{i,j})=d(\tr\{\mathbf{P}^{H}\mathbf{H}^{H}\mathbf{HPe}_{i,j}\mathbf{e}_{i,j}^{H}\})\nonumber \\
\stackrel{(a)}{=}\tr\{(\mathbf{H}^{H}\mathbf{HPe}_{i,j}\mathbf{e}_{i,j}^{H})^{T}d\mathbf{P}^{*}+\mathbf{e}_{i,j}\mathbf{e}_{i,j}^{H}\mathbf{P}^{H}\mathbf{H}^{H}\mathbf{H}d\mathbf{P}\}
\end{gather}
where $(a)$ is obtained from \cite[Table 4.3]{Are2011}. Similarly,
we have
\begin{equation}
\nabla_{\mathbf{P}}\Phi_{i,j}(\boldsymbol{\theta},\mathbf{P})=\mathbf{H}^{H}\mathbf{HPe}_{i,j}\mathbf{e}_{i,j}^{H}.\label{eq:P_grad_2}
\end{equation}
Substituting \eqref{eq:P_grad_2} into \eqref{eq:P_grad_1}, we obtain
\eqref{eq:grad_P}.

\section{Connection Between \eqref{eq:phaseconstraint} and \eqref{eq:relaxphase}
\label{sec:Relax_app}}

Let $\boldsymbol{\theta}^{\ast}$ an the optimal solution to \eqref{eq:rewrite}
and suppose that there is some $l$ such that $\left|\theta_{l}^{\ast}\right|<1$.
Note that we can write $\Phi_{i,j}(\boldsymbol{\theta}^{\ast},\mathbf{P})=\tr\Bigl(\bigl(\mathbf{H}_{\mathrm{DIR}}^{H}+\mathbf{H}_{1}^{H}\mathbf{F}^{H}(\boldsymbol{\theta}^{\ast})\mathbf{H}_{2}^{H}\bigr)\bigl(\mathbf{H}_{\mathrm{DIR}}+\mathbf{H}_{2}\mathbf{F}(\boldsymbol{\theta}^{\ast})\mathbf{H}_{1}\bigr)\mathbf{P}\mathbf{e}_{ij}\mathbf{e}_{ij}^{H}\mathbf{P}^{H}\Bigr)$
as
\[
\Phi_{i,j}(\boldsymbol{\theta}^{\ast},\mathbf{P})=a_{i,j}\left|\theta_{l}^{\ast}\right|^{2}+2\Re\bigl(b_{i,j}\theta_{l}^{\ast}\bigr)+c_{i,j}
\]
where $a_{i,j}\geq0$, $b_{i,j}$ and $c_{i,j}$ are the resulting
constants obtained by rearranging $\Phi_{i,j}(\boldsymbol{\theta}^{\ast},\mathbf{P})$
with respect to $\theta_{l}^{\ast}$. Now consider the point $\tilde{\theta}_{l}=-\theta_{l}^{\ast}$
which is also feasible to \eqref{eq:relaxphase}. Also, we define
$\tilde{\boldsymbol{\theta}}=[\tilde{\theta}_{1},\tilde{\theta}_{2},\ldots,\tilde{\theta}_{N_{\mathrm{ris}}}]^{T}\in\mathbb{C}^{N_{\mathrm{ris}}\times1}$
such that $\tilde{\theta}_{k}=\theta_{k}^{\ast}$ for $k=1,\dots,N_{\mathrm{ris}}$
and $k\neq l$, and $\tilde{\theta}_{l}=-\theta_{l}^{\ast}$. It is
easy to see that $\Phi_{i,j}(\tilde{\boldsymbol{\theta}},\mathbf{P})=a_{i,j}\left|\theta_{l}^{\ast}\right|^{2}-2\Re\bigl(b_{i,j}\theta_{l}^{\ast}\bigr)+c_{i,j}$.
Thus we can conclude that $2\Re\bigl(b_{i,j}\theta_{l}^{\ast}\bigr)\geq0$,
otherwise $\Phi_{i,j}(\tilde{\boldsymbol{\theta}},\mathbf{P})>\Phi_{i,j}(\boldsymbol{\theta}^{\ast},\mathbf{P})$,
and thus, a strictly smaller objective can be obtained. Next, since
$2\Re\bigl(b_{i,j}\theta_{l}^{\ast}\bigr)\geq0$ we immediately have
that $\Phi_{i,j}(\alpha\theta_{l}^{\ast},\mathbf{P})=\alpha^{2}a_{i,j}\left|\theta_{l}^{\ast}\right|^{2}+\alpha2\Re\bigl(b_{i,j}\theta_{l}^{\ast}\bigr)+c_{i,j}\geq a_{i,j}\left|\theta_{l}^{\ast}\right|^{2}+2\Re\bigl(b_{i,j}\theta_{l}^{\ast}\bigr)+c_{i,j}=\Phi_{i,j}(\theta_{l}^{\ast},\mathbf{P})$
for any $\alpha>1$. That means, if $\left|\theta_{l}^{\ast}\right|<1$
then we can always scale $\theta_{l}^{\ast}$ up to achieve a better
objective.\bibliographystyle{IEEEtran}
\bibliography{IEEEabrv,Cut-off_rate_RIS_final}

\end{document}

%% file: acro.tex
\acrodef{AO}{alternating optimization}
\acrodef{ASP}{antenna separation product}
\acrodef{AWGN}{additive white Gaussian noise}
\acrodef{BEP}{bit error probability}
\acrodef{BER}{bit error rate}
\acrodef{BF-MIMO}[BF\mbox{-}MIMO]{beamforming MIMO}
\acrodef{BF}{beamforming}
\acrodef{bpcu}{bits per channel use}
\acrodef{CP}{cyclic prefix}
\acrodef{CR}{cutoff rate}
\acrodef{CSI}{channel state information}
\acrodef{CSIR}{channel state information at RX}
\acrodef{SSK}{space shift keying}
\acrodef{CSIT}{channel state information at TX}
\acrodef{DCMC}{discrete\mbox{-}input continuous\mbox{-}output memoryless channel}
\acrodef{DFT}{discrete Fourier transform}
\acrodef{DL-TR-GSM}{dual-layered transmit-receive \acl{GSM}}
\acrodef{DLT}{dual-layered transmission}
\acrodef{EGC}{equal gain combining}
\acrodef{EM}{electromagnetic}
\acrodef{FSPL}{free space path loss}
\acrodef{FFT}{fast Fourier transform}
\acrodef{FDE}{frequency domain equalization}
\acrodef{GRSM}{generalized \acl{RSM}}
\acrodef{GSM}{generalized \acl{SM}}
\acrodef{IFFT}{invserse fast Fourier transform}
\acrodef{ICI}{inter-channel interference}
\acrodef{iid}[i.i.d.]{independent and identically distributed}
\acrodef{IQ}{in\mbox{-}phase and quadrature}
\acrodef{ISI}{intersymbol interference}
\acrodef{ISI-free}[ISI\mbox{-}free]{intersymbol interference free}
\acrodef{LIS}{large intelligent surface}
\acrodef{LOS}{line\mbox{-}of\mbox{-}sight}
\acrodef{mmWave}{millimeter-wave}
\acrodef{MIMO}{multiple\mbox{-}input multiple\mbox{-}output}
\acrodef{MISO}{multiple\mbox{-}input single\mbox{-}output}
\acrodef{MI}{mutual information}
\acrodef{ML}{maximum likelihood}
\acrodef{MRC}{maximal ratio combining}
\acrodef{MMSE}{minimum mean square error}
\acrodef{MU-TR-GSM}{multiuser transmit-receive  \acl{GSM} }
\acrodef{NCSIT}{no channel state information at TX}
\acrodef{NLOS}{non\mbox{-}\acs{LOS}} 
\acrodef{NOMA}{non-orthogonal multiple access}
\acrodef{OFDM}{orthogonal frequency division multiplexing}
\acrodef{OFDMA}{orthogonal frequency division multiple access}
\acrodef{PA}{power amplifier}
\acrodef{PAE}{power added efficiency}
\acrodef{PAPR}{peak\mbox{-}to\mbox{-}average power ratio}
\acrodef{PDF}{probability density function}
\acrodef{PEP}{pairwise error probability}
\acrodefplural{PEP}{pairwise error probabilities}
\acrodef{PGM}{projected gradient method}
\acrodef{APGM}{accelerated projected gradient method}
\acrodef{PMP}{probability mass function}
\acrodef{PSM}{precoding-aided spatial modulation}
\acrodef{QSM}{quadrature spatial modulation}
\acrodef{RC}{reorganization computation}
\acrodef{RIS}{reconfigurable intelligent surface}
\acrodef{RSM}{receive spatial modulation}
\acrodef{RX}{receiver}
\acrodef{SCA}{successive convex approximation}
\acrodef{SEP}{symbol error probability}
\acrodef{SER}{symbol error rate}
\acrodef{SINR}{signal-to-interference-plus-noise ratio}
\acrodef{SISO}{single-input single-output}
\acrodef{SM}{spatial modulation}
\acrodef{SMX-MIMO}[SMX\mbox{-}MIMO]{spatial multiplexing MIMO}
\acrodef{SMX}{spatial multiplexing}
\acrodef{SNR}{signal-to-noise ratio}
\acrodef{SC}{single carrier}
\acrodef{SVD}{singular value decomposition}
\acrodef{SPST}{single pole single-throw}
\acrodef{SU}{secondary user}
\acrodef{TDE}{time domain equalization}
\acrodef{TX}{transmitter}
\acrodef{ULA}{uniform linear array}
\acrodef{URA}{uniform rectangular array}
\acrodef{VGA}{variable gain amplifier}
\acrodef{ZF}{zero-forcing}
\acrodef{ZMCG}{zero-mean complex Gaussian}

%% file: figures_scaled/plot_8_2_dir_sc_V1.tex
\begin{tikzpicture}[every node/.style={scale=0.80}]
\pgfplotsset{width=0.4\columnwidth,height =0.45\columnwidth, compat=1.6}
\begin{axis}[%
scale only axis,
separate axis lines,
every outer x axis line/.append style={black},
every x tick label/.append style={font=\color{black}},
xmin=0,
xmax=20,
xlabel={Iteration},
every outer y axis line/.append style={black},
every y tick label/.append style={font=\color{black}},
ymin=0.845553924352954,
ymax=3.8,
ylabel={$R_0$, MI [bpcu]},
axis background/.style={fill=white},
legend style={at={(1,0.35)},anchor=south east,legend cell align=left,align=left,draw=black,inner sep=0pt,row sep=-2pt,font=\footnotesize}
]
\addplot [color=blue,solid,line width=1.0pt]
  table[row sep=crcr]{%
1	0.845553924352954\\
2	2.5453314318706\\
3	3.04566263044777\\
4	3.26947834289609\\
5	3.32129381114271\\
6	3.3535771039231\\
7	3.36818283857915\\
8	3.38867059981681\\
9	3.39530585430329\\
10	3.39997210052176\\
11	3.40396216750198\\
12	3.40829642199885\\
13	3.41252860216425\\
14	3.41674854771484\\
15	3.4190759743707\\
16	3.42059695280019\\
17	3.42198267565589\\
18	3.42327219227683\\
19	3.42447580865788\\
20	3.42560934854832\\
21	3.42667681983195\\
22	3.42768550548451\\
23	3.42863761688067\\
24	3.42953700034191\\
25	3.43038564523674\\
26	3.43118622106833\\
27	3.43194050730529\\
28	3.43264986684839\\
29	3.43331485354853\\
30	3.43393635396142\\
31	3.43451655899201\\
32	3.43505875100799\\
33	3.43556616032924\\
34	3.43604146782554\\
35	3.4364869361355\\
36	3.43690459606187\\
37	3.43729633010354\\
38	3.43766389685122\\
39	3.43800894069554\\
40	3.43833299618092\\
41	3.43863749367151\\
42	3.43892376420197\\
43	3.43919304550076\\
44	3.43944648748926\\
45	3.43968515826187\\
46	3.43991004958379\\
47	3.4401220824459\\
48	3.44032211216618\\
49	3.4405109333299\\
50	3.44068928431009\\
51	3.44085785152602\\
52	3.44101727331245\\
53	3.44116814348407\\
54	3.44131101453753\\
55	3.441446400547\\
56	3.44157477975042\\
57	3.4416965968955\\
58	3.44181226539685\\
59	3.4419221693892\\
60	3.44202666573225\\
61	3.44212608600311\\
62	3.44222073846322\\
63	3.44231090996114\\
64	3.44239686771866\\
65	3.44247886096275\\
66	3.44255712238655\\
67	3.44263186944551\\
68	3.44270330550684\\
69	3.44277162087543\\
70	3.44283699371768\\
71	3.44289959090072\\
};
\addlegendentry{$R_0$, PGM};

\addplot [color=blue,dotted,line width=1.0pt]
  table[row sep=crcr]{%
1	1.30194404309574\\
2	3.00019274304825\\
3	3.39135224410481\\
4	3.58193270729633\\
5	3.62891540154498\\
6	3.6553070729484\\
7	3.66476498062283\\
8	3.68717542758516\\
9	3.68957959072663\\
10	3.69257039603056\\
11	3.69697693302775\\
12	3.69793040704751\\
13	3.70420210009988\\
14	3.70770263877038\\
15	3.70695659213457\\
16	3.70837668781701\\
17	3.70910725230524\\
18	3.71124149100016\\
19	3.71365788986061\\
20	3.7131649857501\\
21	3.71282073654959\\
22	3.71333451329876\\
23	3.71416550817594\\
24	3.71539587210796\\
25	3.71558457399283\\
26	3.71575059567755\\
27	3.71644548182544\\
28	3.71744261642915\\
29	3.72036721206598\\
30	3.72015361075503\\
31	3.71998790380487\\
32	3.71940016646768\\
33	3.71787172088112\\
34	3.72001716949111\\
35	3.71898793980036\\
36	3.72033927046414\\
37	3.7197829506316\\
38	3.72205415526979\\
39	3.72306835435382\\
40	3.72323678876701\\
41	3.72440584976821\\
42	3.72252885152767\\
43	3.72459147282321\\
44	3.72157730900949\\
45	3.72181170604691\\
46	3.72144411373325\\
47	3.72145522181689\\
48	3.72440171832718\\
49	3.72001009321379\\
50	3.72013807521024\\
51	3.71937202455363\\
52	3.72389667829505\\
53	3.72447507285214\\
54	3.72329737515514\\
55	3.72240963300493\\
56	3.7200963783827\\
57	3.72171053664895\\
58	3.72100866755369\\
59	3.72525264439416\\
60	3.72327857425271\\
61	3.72276747059696\\
62	3.72334584029343\\
63	3.72232783177031\\
64	3.72568049630437\\
65	3.72161504216269\\
66	3.72621293922293\\
67	3.7255721401938\\
68	3.7252618463366\\
69	3.72558260775549\\
70	3.7238055409687\\
71	3.72550827226874\\
};
\addlegendentry{MI, PGM};

\addplot [color=red,solid,line width=1.0pt]
  table[row sep=crcr]{%
1	0.93691572681466\\
2	2.57220970450482\\
3	3.02969689950812\\
4	3.25564551507821\\
5	3.31164544685331\\
6	3.33227330792722\\
7	3.34245065317332\\
8	3.34915114424818\\
9	3.35393801865373\\
10	3.3576937802949\\
11	3.36089965633507\\
12	3.3637599161625\\
13	3.36635997149988\\
14	3.36874584717826\\
15	3.3709550153878\\
16	3.37302444359968\\
17	3.37498908086226\\
18	3.37687897461374\\
19	3.3787175833916\\
20	3.38052119483982\\
21	3.38229871979018\\
22	3.3840519025707\\
23	3.38577588673823\\
24	3.38746044546472\\
25	3.38909193013574\\
26	3.39065693352508\\
27	3.39214138129731\\
28	3.39353103653984\\
29	3.39481853102276\\
30	3.39600046087551\\
31	3.39707726005821\\
32	3.39805272501554\\
33	3.39893395167155\\
34	3.39973356150136\\
35	3.40045602836218\\
36	3.40110254319813\\
37	3.40168269027722\\
38	3.40220454737902\\
39	3.40267508704833\\
40	3.40310049019403\\
41	3.40348615754165\\
42	3.40383679876225\\
43	3.40415647874579\\
44	3.40444869075333\\
45	3.40471643914453\\
46	3.40496232226411\\
47	3.40518861451546\\
48	3.40539732745251\\
49	3.40559025436911\\
50	3.40576902255663\\
51	3.40593508535534\\
52	3.40608976796664\\
53	3.40623424425198\\
54	3.40636957205194\\
55	3.40649668209751\\
56	3.40661639173293\\
57	3.40672943610244\\
58	3.40683644462923\\
59	3.4069379821668\\
60	3.40703453718365\\
61	3.40712654303413\\
62	3.40721438252773\\
63	3.40729838958551\\
64	3.40737886380925\\
65	3.40745606915137\\
66	3.40753024255218\\
67	3.40760158928036\\
68	3.40767029692801\\
69	3.40773652969884\\
70	3.40780044255219\\
71	3.40786216705444\\
};
\addlegendentry{$R_0$, SCA};

\addplot [color=red,dotted,line width=1.0pt]
  table[row sep=crcr]{%
1	1.41988591111634\\
2	3.00494016172516\\
3	3.38450354482623\\
4	3.58195660001536\\
5	3.62735944992751\\
6	3.64401561330677\\
7	3.65657533665304\\
8	3.65959509413147\\
9	3.66406917285318\\
10	3.66292910118571\\
11	3.66779630640662\\
12	3.66992850371214\\
13	3.67256092206266\\
14	3.67454681233237\\
15	3.67751302536752\\
16	3.67675108297107\\
17	3.67780242656514\\
18	3.67813365896602\\
19	3.67977208456653\\
20	3.68164779365599\\
21	3.68517189735365\\
22	3.68360459419772\\
23	3.68392502781845\\
24	3.68653517566263\\
25	3.68722344711905\\
26	3.68624342171994\\
27	3.68789940418795\\
28	3.69210160321168\\
29	3.69054333909444\\
30	3.6906214049139\\
31	3.69135576769914\\
32	3.69221964753875\\
33	3.69155552637263\\
34	3.69691833014036\\
35	3.69327493285635\\
36	3.69163779269082\\
37	3.69833389620651\\
38	3.6959338478345\\
39	3.69719223065329\\
40	3.69510428121425\\
41	3.69945311097563\\
42	3.69763789953323\\
43	3.6982181345264\\
44	3.7004618999016\\
45	3.69735362499156\\
46	3.69880033249713\\
47	3.70036122390292\\
48	3.69942962208324\\
49	3.69669556836218\\
50	3.69831478111045\\
51	3.69537245456992\\
52	3.69610854006569\\
53	3.6976175260626\\
54	3.69909426484693\\
55	3.69996460782561\\
56	3.70118154860741\\
57	3.69581775580351\\
58	3.70027172597257\\
59	3.70073890380852\\
60	3.69866512251381\\
61	3.69962489065883\\
62	3.69729473776487\\
63	3.69953701202653\\
64	3.70171575154287\\
65	3.7006846077583\\
66	3.70384788980157\\
67	3.7004802652882\\
68	3.70122649107138\\
69	3.69761840960432\\
70	3.69666460570585\\
71	3.70293993780069\\
};
\addlegendentry{MI, SCA};

\addplot [color=green,solid,line width=1.0pt,mark size=2pt,mark=x,mark options={solid}]
  table[row sep=crcr]{%
1	0.877580587383905\\
2	1.2937203197227\\
3	1.30373764174731\\
4	1.30637270311322\\
5	1.30746207202656\\
6	1.30847190942383\\
7	1.30896424779131\\
8	1.30910308075706\\
9	1.30917532001215\\
10	1.30921166670911\\
11	1.30923887335242\\
12	1.30925811963936\\
13	1.30927124450894\\
14	1.30928011204425\\
15	1.30928622069479\\
16	1.30929060797218\\
17	1.30929392544959\\
18	1.30929655935059\\
19	1.30929655935059\\
20	1.30929655935059\\
21	1.30929655935059\\
22	1.30929655935059\\
23	1.30929655935059\\
24	1.30929655935059\\
25	1.30929655935059\\
26	1.30929655935059\\
27	1.30929655935059\\
28	1.30929655935059\\
29	1.30929655935059\\
30	1.30929655935059\\
31	1.30929655935059\\
32	1.30929655935059\\
33	1.30929655935059\\
34	1.30929655935059\\
35	1.30929655935059\\
36	1.30929655935059\\
37	1.30929655935059\\
38	1.30929655935059\\
39	1.30929655935059\\
40	1.30929655935059\\
41	1.30929655935059\\
};
\addlegendentry{$R_0$, PGM, no RIS};

\addplot [color=green,dotted,line width=1.0pt,mark size=2pt,mark=x,mark options={solid}]
  table[row sep=crcr]{%
1	1.34383910485268\\
2	1.83205725484852\\
3	1.84148552480936\\
4	1.84238367548191\\
5	1.8430609737965\\
6	1.84415731017079\\
7	1.84436637495188\\
8	1.84492850464255\\
9	1.84524644558875\\
10	1.84508665607173\\
11	1.8452051087093\\
12	1.84572390015402\\
13	1.84586331790699\\
14	1.84585648615636\\
15	1.84579411525496\\
16	1.84585398091191\\
17	1.84602839571189\\
18	1.84607114521828\\
19	1.84607114521828\\
20	1.84607114521828\\
21	1.84607114521828\\
22	1.84607114521828\\
23	1.84607114521828\\
24	1.84607114521828\\
25	1.84607114521828\\
26	1.84607114521828\\
27	1.84607114521828\\
28	1.84607114521828\\
29	1.84607114521828\\
30	1.84607114521828\\
31	1.84607114521828\\
32	1.84607114521828\\
33	1.84607114521828\\
34	1.84607114521828\\
35	1.84607114521828\\
36	1.84607114521828\\
37	1.84607114521828\\
38	1.84607114521828\\
39	1.84607114521828\\
40	1.84607114521828\\
41	1.84607114521828\\
};
\addlegendentry{MI, PGM, no RIS};

\addplot [color=mycolor1,solid,line width=1.0pt]
  table[row sep=crcr]{%
1	0.885853153485951\\
2	1.29340790747348\\
3	1.30334258912439\\
4	1.30529333568079\\
5	1.3062422049014\\
6	1.30694016335838\\
7	1.30738447020141\\
8	1.30767680557311\\
9	1.30778314530724\\
10	1.30786670907175\\
11	1.30793840360468\\
12	1.30800666214695\\
13	1.30807768824877\\
14	1.30815577708182\\
15	1.30816392685157\\
16	1.30816392685157\\
17	1.30816392685157\\
18	1.30816392685157\\
19	1.30816392685157\\
20	1.30816392685157\\
21	1.30816392685157\\
22	1.30816392685157\\
23	1.30816392685157\\
24	1.30816392685157\\
25	1.30816392685157\\
26	1.30816392685157\\
27	1.30816392685157\\
28	1.30816392685157\\
29	1.30816392685157\\
30	1.30816392685157\\
31	1.30816392685157\\
32	1.30816392685157\\
33	1.30816392685157\\
34	1.30816392685157\\
35	1.30816392685157\\
36	1.30816392685157\\
37	1.30816392685157\\
38	1.30816392685157\\
39	1.30816392685157\\
40	1.30816392685157\\
41	1.30816392685157\\
};
\addlegendentry{$R_0$, SCA, no RIS};

\addplot [color=mycolor1,dotted,line width=1.0pt]
  table[row sep=crcr]{%
1	1.35232801587445\\
2	1.83909306365949\\
3	1.84790313881095\\
4	1.85003267118993\\
5	1.85148891248141\\
6	1.85238903789076\\
7	1.85504616773871\\
8	1.85558897853353\\
9	1.85568515241281\\
10	1.85569186078145\\
11	1.85593928956301\\
12	1.85560940775765\\
13	1.85586158503344\\
14	1.856427525151\\
15	1.85651423857173\\
16	1.85651423857173\\
17	1.85651423857173\\
18	1.85651423857173\\
19	1.85651423857173\\
20	1.85651423857173\\
21	1.85651423857173\\
22	1.85651423857173\\
23	1.85651423857173\\
24	1.85651423857173\\
25	1.85651423857173\\
26	1.85651423857173\\
27	1.85651423857173\\
28	1.85651423857173\\
29	1.85651423857173\\
30	1.85651423857173\\
31	1.85651423857173\\
32	1.85651423857173\\
33	1.85651423857173\\
34	1.85651423857173\\
35	1.85651423857173\\
36	1.85651423857173\\
37	1.85651423857173\\
38	1.85651423857173\\
39	1.85651423857173\\
40	1.85651423857173\\
41	1.85651423857173\\
};
\addlegendentry{MI, SCA, no RIS};

\end{axis}
\end{tikzpicture}%

%% file: figures_scaled/plot_8_2_nondir_sc.tex
%


\begin{tikzpicture}[every node/.style={scale=0.80}]
\pgfplotsset{width=0.4\columnwidth,height =0.45\columnwidth, compat=1.6}

\begin{axis}[%
scale only axis,
separate axis lines,
every outer x axis line/.append style={black},
every x tick label/.append style={font=\color{black}},
xmin=0,
xmax=20,
xlabel={Iteration},
every outer y axis line/.append style={black},
every y tick label/.append style={font=\color{black}},
ymin=0,
ymax=3.8,
ylabel={$R_0$, MI [bpcu]},
axis background/.style={fill=white},
legend style={at={(0.97,0.03)},anchor=south east,legend cell align=left,align=left,draw=black,inner sep=0pt,row sep=-2pt,font=\footnotesize}
]
\addplot [color=blue,solid,line width=1.0pt]
  table[row sep=crcr]{%
1	0.0167238676261835\\
2	1.70962322391565\\
3	2.92210480584456\\
4	3.03193928805954\\
5	3.05686754872683\\
6	3.06503733570197\\
7	3.06836871725404\\
8	3.06984562774151\\
9	3.07115406067323\\
10	3.071661159065\\
11	3.07211870714976\\
12	3.07226874775976\\
13	3.07236274190764\\
14	3.07243078339291\\
15	3.0725116578375\\
16	3.07255203373518\\
17	3.07258202041436\\
18	3.07260533410647\\
19	3.07262369660235\\
20	3.07263831267201\\
21	3.07265006167614\\
22	3.07265959554568\\
23	3.07266740245294\\
24	3.07267385221779\\
25	3.0726792275346\\
26	3.07268374670253\\
27	3.07268757964022\\
28	3.07269085966148\\
29	3.07269369196669\\
30	3.07269615997114\\
31	3.07269832999709\\
32	3.0727002548475\\
33	3.07270197654377\\
34	3.07270352846976\\
35	3.07270493706872\\
36	3.07270622320891\\
37	3.07270740329313\\
38	3.07270849016956\\
39	3.07270949388324\\
40	3.07271042229786\\
41	3.07271128160939\\
42	3.072712076768\\
43	3.07271281182095\\
44	3.07271349018683\\
45	3.07271411486999\\
46	3.07271468862294\\
47	3.07271521406399\\
48	3.07271569375708\\
49	3.07271613026042\\
50	3.07271652615018\\
51	3.07271688402513\\
52	3.07271720649753\\
53	3.07271749617505\\
54	3.07271775563755\\
55	3.07271798741208\\
56	3.0727181939485\\
57	3.07271837759749\\
58	3.07271854059218\\
59	3.07271868503379\\
60	3.07271881288174\\
61	3.07271892594781\\
62	3.07271902589421\\
63	3.07271911423482\\
64	3.07271919233931\\
65	3.07271926143922\\
66	3.07271932263574\\
67	3.0727193769085\\
68	3.07271942512492\\
69	3.07271946804994\\
70	3.07271950635558\\
71	3.07271954063025\\
};
\addlegendentry{$R_0$, PGM};

\addplot [color=blue,dotted,line width=1.0pt]
  table[row sep=crcr]{%
1	0.0332240904913683\\
2	2.20886329427005\\
3	3.30633365018273\\
4	3.41929724430214\\
5	3.4487189240278\\
6	3.45455627987058\\
7	3.4633973645969\\
8	3.46378954367598\\
9	3.46651176300832\\
10	3.46787453605922\\
11	3.46722788021659\\
12	3.46951255239921\\
13	3.46928454613378\\
14	3.46463414622926\\
15	3.46867088566247\\
16	3.46918536903802\\
17	3.46888979307838\\
18	3.46753630320335\\
19	3.46651100798365\\
20	3.46910438049192\\
21	3.46622412846115\\
22	3.46736068198421\\
23	3.46926507537258\\
24	3.46846666576853\\
25	3.46881175173124\\
26	3.46694164014014\\
27	3.46906660781673\\
28	3.46673270199629\\
29	3.46535397201565\\
30	3.46834817528836\\
31	3.46905045174142\\
32	3.46917843093141\\
33	3.46526293187317\\
34	3.46581670700809\\
35	3.46919633575177\\
36	3.46727201588144\\
37	3.46635470196659\\
38	3.46709701884098\\
39	3.46637511810076\\
40	3.46764105263584\\
41	3.46963303411783\\
42	3.46705531768852\\
43	3.46777647206405\\
44	3.46826750849361\\
45	3.46843529001795\\
46	3.46683820206769\\
47	3.4678589412695\\
48	3.46922875481188\\
49	3.46705088156169\\
50	3.46666530413378\\
51	3.46804785743333\\
52	3.46915259260562\\
53	3.46917515940075\\
54	3.46986416641898\\
55	3.46827723896172\\
56	3.46890790340879\\
57	3.46473837246784\\
58	3.46684727685768\\
59	3.4678115895155\\
60	3.46900085099948\\
61	3.46995823128268\\
62	3.46760018541456\\
63	3.47129528300726\\
64	3.46851372173357\\
65	3.46858624366796\\
66	3.46598950658652\\
67	3.4674200360484\\
68	3.46493686685602\\
69	3.46332768882312\\
70	3.46456559463394\\
71	3.4673538581039\\
};
\addlegendentry{MI, PGM};

\addplot [color=red,solid,line width=1.0pt]
  table[row sep=crcr]{%
1	0.0162448100416202\\
2	2.74339548201864\\
3	3.02610810240511\\
4	3.05403164565772\\
5	3.06294532584725\\
6	3.06709827510565\\
7	3.06984753932191\\
8	3.07212253309477\\
9	3.07407892313712\\
10	3.0755442996076\\
11	3.07651624428122\\
12	3.07715344551532\\
13	3.07758478496343\\
14	3.07788516138001\\
15	3.07809871337172\\
16	3.07825331351202\\
17	3.07836725519251\\
18	3.07845268102845\\
19	3.07851781895864\\
20	3.07856827364902\\
21	3.0786079378596\\
22	3.07863953150352\\
23	3.07866500438792\\
24	3.07868577631036\\
25	3.07870286771891\\
26	3.07871704971685\\
27	3.07872891897055\\
28	3.07873890690777\\
29	3.07874736012117\\
30	3.07875456618887\\
31	3.07876072181695\\
32	3.07876600558983\\
33	3.07877055389211\\
34	3.07877448105278\\
35	3.07877788408393\\
36	3.0787808373894\\
37	3.07878340473943\\
38	3.07878564921505\\
39	3.07878760384395\\
40	3.07878931372326\\
41	3.07879080829762\\
42	3.07879211778196\\
43	3.07879326436868\\
44	3.07879427951256\\
45	3.07879516724293\\
46	3.07879595402913\\
47	3.07879664500546\\
48	3.078797254443\\
49	3.0787977921229\\
50	3.07879826691869\\
51	3.0787986875992\\
52	3.07879906001452\\
53	3.07879938988579\\
54	3.07879968017257\\
55	3.07879994029181\\
56	3.07880017545753\\
57	3.07880037909754\\
58	3.07880056703675\\
59	3.07880073345158\\
60	3.07880087970413\\
61	3.07880101211\\
62	3.07880113144392\\
63	3.07880123856404\\
64	3.07880133466061\\
65	3.07880142276103\\
66	3.07880150306854\\
67	3.07880157638812\\
68	3.07880164151769\\
69	3.07880170127512\\
70	3.078801755059\\
71	3.07880180672159\\
};
\addlegendentry{$R_0$, SCA};

\addplot [color=red,dotted,line width=1.0pt]
  table[row sep=crcr]{%
1	0.0309818836920726\\
2	3.15503850321577\\
3	3.41198878131481\\
4	3.44322215664018\\
5	3.45610864356314\\
6	3.45885443713998\\
7	3.46257940690626\\
8	3.464153886408\\
9	3.4694738714257\\
10	3.4670817477403\\
11	3.47394888458248\\
12	3.47506325820713\\
13	3.47103116469165\\
14	3.47559076250773\\
15	3.47389118556842\\
16	3.47670727056289\\
17	3.47736879763851\\
18	3.47722745684526\\
19	3.47712984153978\\
20	3.46989497865532\\
21	3.47357908703421\\
22	3.47935045403776\\
23	3.47298470508598\\
24	3.47259577046854\\
25	3.46952715345885\\
26	3.47520079812892\\
27	3.47779309147211\\
28	3.47565248747976\\
29	3.47490920251567\\
30	3.47563388049273\\
31	3.47603050766338\\
32	3.47695449138634\\
33	3.47380168281731\\
34	3.47138443759167\\
35	3.47180722779088\\
36	3.47452916273077\\
37	3.48165802161335\\
38	3.47688540658066\\
39	3.47290530535737\\
40	3.47186723411213\\
41	3.47286754930639\\
42	3.47521454007916\\
43	3.47239305838181\\
44	3.47454599073891\\
45	3.47694098279408\\
46	3.47609246501746\\
47	3.47486571875341\\
48	3.47450082893056\\
49	3.47518098255342\\
50	3.47414737205457\\
51	3.47496381665767\\
52	3.47312134974045\\
53	3.47482262942662\\
54	3.47651441807594\\
55	3.4770473624673\\
56	3.47262647421405\\
57	3.47240036422532\\
58	3.47537553342222\\
59	3.47463300367011\\
60	3.46918266951241\\
61	3.47515062986588\\
62	3.47301484202515\\
63	3.47445429127107\\
64	3.47117729925815\\
65	3.47176996693766\\
66	3.47149314461679\\
67	3.48108186282443\\
68	3.47798929943758\\
69	3.47364395291529\\
70	3.47697508607878\\
71	3.47357141356744\\
};
\addlegendentry{MI, SCA};

\end{axis}
\end{tikzpicture}%

%% file: figures_scaled/dis_vs_cont_dir_sc.tex
%


\begin{tikzpicture}[every node/.style={scale=0.80}]

\pgfplotsset{width=0.4\columnwidth,height =0.45\columnwidth, compat=1.6,
legend image code/.code={
	\draw[mark repeat=1,mark phase=1]
	plot coordinates {
		(0cm,0cm)
		(0cm,0cm)        
		(0.3cm,0cm)         
	};%
}
}
\begin{axis}[%
scale only axis,
separate axis lines,
every outer x axis line/.append style={black},
every x tick label/.append style={font=\color{black}},
xmin=0,
xmax=20,
xlabel={Iteration},
every outer y axis line/.append style={black},
every y tick label/.append style={font=\color{black}},
ymin=1,
ymax=4.5,
ylabel={MI [bpcu]},
axis background/.style={fill=white},
legend style={at={(1,0.0)},anchor=south east,legend cell align=left,align=left,draw=black,inner sep=1pt,row sep=-2pt,font=\footnotesize}
]
\addplot [color=blue,solid,line width=1.0pt]
  table[row sep=crcr]{%
1	1.30194404309574\\
2	3.00019274304825\\
3	3.39135224410481\\
4	3.58193270729633\\
5	3.62891540154498\\
6	3.6553070729484\\
7	3.66476498062283\\
8	3.68717542758516\\
9	3.68957959072663\\
10	3.69257039603056\\
11	3.69697693302775\\
12	3.69793040704751\\
13	3.70420210009988\\
14	3.70770263877038\\
15	3.70695659213457\\
16	3.70837668781701\\
17	3.70910725230524\\
18	3.71124149100016\\
19	3.71365788986061\\
20	3.7131649857501\\
21	3.71282073654959\\
22	3.71333451329876\\
23	3.71416550817594\\
24	3.71539587210796\\
25	3.71558457399283\\
26	3.71575059567755\\
27	3.71644548182544\\
28	3.71744261642915\\
29	3.72036721206598\\
30	3.72015361075503\\
31	3.71998790380487\\
32	3.71940016646768\\
33	3.71787172088112\\
34	3.72001716949111\\
35	3.71898793980036\\
36	3.72033927046414\\
37	3.7197829506316\\
38	3.72205415526979\\
39	3.72306835435382\\
40	3.72323678876701\\
41	3.72440584976821\\
42	3.72252885152767\\
43	3.72459147282321\\
44	3.72157730900949\\
45	3.72181170604691\\
46	3.72144411373325\\
47	3.72145522181689\\
48	3.72440171832718\\
49	3.72001009321379\\
50	3.72013807521024\\
51	3.71937202455363\\
52	3.72389667829505\\
53	3.72447507285214\\
54	3.72329737515514\\
55	3.72240963300493\\
56	3.7200963783827\\
57	3.72171053664895\\
58	3.72100866755369\\
59	3.72525264439416\\
60	3.72327857425271\\
61	3.72276747059696\\
62	3.72334584029343\\
63	3.72232783177031\\
64	3.72568049630437\\
65	3.72161504216269\\
66	3.72621293922293\\
67	3.7255721401938\\
68	3.7252618463366\\
69	3.72558260775549\\
70	3.7238055409687\\
71	3.72550827226874\\
};
\addlegendentry{$M=4$, CR-based opt.};

\addplot [color=blue,dotted,line width=1.0pt]
  table[row sep=crcr]{%
1	1.32198894470111\\
2	3.21110838484402\\
3	3.91796783254211\\
4	4.13894689891465\\
5	4.1749297172361\\
6	4.18065178879545\\
7	4.17837396744256\\
8	4.17846381939975\\
9	4.17833280093284\\
10	4.1835784619812\\
11	4.18459991179288\\
12	4.1885910740921\\
13	4.19222328893514\\
14	4.1994212030548\\
15	4.20076880933027\\
16	4.20559143513352\\
17	4.21083731420373\\
18	4.21374412684017\\
19	4.21481946126354\\
20	4.21873885660599\\
21	4.21986930074819\\
22	4.22237469185851\\
23	4.22558632417318\\
24	4.22751184341384\\
25	4.22827871585902\\
26	4.2297907985588\\
27	4.2326204545252\\
28	4.23288962997345\\
29	4.23535579934347\\
30	4.23761232661745\\
31	4.23843662589751\\
32	4.23710875412703\\
33	4.23999408737752\\
34	4.24075334136108\\
35	4.24145893070722\\
36	4.24322313057146\\
37	4.24310484477357\\
38	4.24494376142407\\
39	4.24489581690152\\
40	4.24579676687029\\
41	4.24657360099329\\
42	4.24635030784933\\
43	4.24656306004056\\
44	4.2476050380166\\
45	4.24776595793407\\
46	4.24799773339685\\
47	4.24921713137239\\
48	4.24755445294065\\
49	4.24782639445429\\
50	4.25031203809545\\
51	4.25108066414792\\
52	4.25010242323548\\
53	4.24916991471511\\
54	4.25135903304825\\
55	4.2503156592177\\
56	4.25020152974391\\
57	4.25100893456076\\
58	4.25177261076559\\
59	4.25201147782514\\
60	4.25174035898582\\
61	4.25227537592659\\
62	4.25162346242829\\
63	4.25166612230482\\
64	4.25257118907378\\
65	4.2524469171128\\
66	4.25104606442736\\
67	4.2530399805623\\
68	4.25283495685907\\
69	4.25155674407751\\
70	4.25029500789105\\
71	4.25181351590924\\
};
\addlegendentry{$M=16$, CR-based opt.};

\addplot [color=red,solid,line width=1.0pt]
  table[row sep=crcr]{%
1	1.3619562743261\\
2	1.89344468229566\\
3	2.0877563780813\\
4	2.16903942779127\\
5	2.28193750952217\\
6	2.24511248837461\\
7	2.24758189195328\\
8	2.23059764927046\\
9	2.22472838523933\\
10	2.22245191281721\\
11	2.21848509722159\\
12	2.22499293630375\\
13	2.2174611187946\\
14	2.222084795391\\
15	2.21293450147664\\
16	2.21434660836442\\
17	2.21074707380726\\
18	2.20932486069969\\
19	2.20816662593115\\
20	2.20651554971226\\
21	2.2066742238904\\
22	2.20454962742523\\
23	2.20404637864853\\
24	2.20324379521588\\
25	2.20238748882044\\
26	2.20110667076926\\
27	2.19979619785726\\
28	2.19987595140276\\
29	2.1982506015457\\
30	2.19925520280523\\
31	2.19712151631425\\
32	2.19837969454613\\
33	2.19713052745268\\
34	2.19741112531387\\
35	2.19680599734449\\
36	2.19773892564819\\
37	2.19642189888982\\
38	2.19680192269731\\
39	2.1961846775327\\
40	2.19707705873599\\
41	2.19606975349459\\
};
\addlegendentry{$M=4$, Gauss-based opt.};

\addplot [color=red,dotted,line width=1.0pt]
  table[row sep=crcr]{%
1	1.32390668380815\\
2	2.35244920503875\\
3	2.64697864802257\\
4	3.2023951023816\\
5	3.42944199627911\\
6	3.55087992847742\\
7	3.60346863928398\\
8	3.64243645525053\\
9	3.68845767235615\\
10	3.71807036487793\\
11	3.74060304437369\\
12	3.75111022880056\\
13	3.76167318863643\\
14	3.76524437068436\\
15	3.7697756768579\\
16	3.77136778136572\\
17	3.77447904709344\\
18	3.77418991840575\\
19	3.77673236610068\\
20	3.77753692302032\\
21	3.77921155646217\\
22	3.77978472416306\\
23	3.7804321734654\\
24	3.78045633750329\\
25	3.78146042923259\\
26	3.78084539876425\\
27	3.78136124848552\\
28	3.78273147233962\\
29	3.78150202042389\\
30	3.78272550761793\\
31	3.78396627218273\\
32	3.78425238392072\\
33	3.7844438063419\\
34	3.78406288753394\\
35	3.78607897948797\\
36	3.78546893530197\\
37	3.7855517816598\\
38	3.78575228207243\\
39	3.78569074429968\\
40	3.78691789031901\\
41	3.78685802661161\\
};
\addlegendentry{$M=16$, Gauss-based opt.};

\addplot [color=green,solid,line width=1.0pt]
  table[row sep=crcr]{%
1	1.41664919129579\\
2	2.5554274529707\\
3	2.91488311950573\\
4	3.46704064256436\\
5	3.82849260676017\\
6	4.0037115873677\\
7	4.08670202753089\\
8	4.14718765555675\\
9	4.19423693898005\\
10	4.22827568976702\\
11	4.26217033456645\\
12	4.2836063272395\\
13	4.29599974464836\\
14	4.30724998357526\\
15	4.31495626045316\\
16	4.32142815677068\\
17	4.32509182330441\\
18	4.32782320755621\\
19	4.32991816007117\\
20	4.33250429021851\\
21	4.33409172077454\\
22	4.33541622956549\\
23	4.33645724306601\\
24	4.33746993125273\\
25	4.33847705484307\\
26	4.33939698853758\\
27	4.34015245166163\\
28	4.34089349788946\\
29	4.34158290676142\\
30	4.34223368151921\\
31	4.34296653496468\\
32	4.34355069084242\\
33	4.34404793854832\\
34	4.34451241956795\\
35	4.34493401861602\\
36	4.3453337713322\\
37	4.34570302019536\\
38	4.346055970224\\
39	4.34638420838508\\
40	4.34669838725372\\
41	4.34699240353118\\
42	4.34727442911705\\
43	4.34753985064587\\
44	4.34779490725808\\
45	4.34803606501705\\
46	4.3482680342705\\
47	4.34848805103927\\
48	4.34869960833846\\
49	4.34890052185995\\
50	4.34909333482383\\
51	4.34927642438421\\
52	4.34945175578808\\
53	4.34961832448431\\
54	4.34977771457296\\
55	4.34993620854412\\
56	4.35008420873277\\
57	4.35022389956455\\
58	4.35035671287464\\
59	4.35048231174249\\
60	4.35060335620893\\
61	4.3507167411357\\
62	4.35086374268648\\
63	4.35098395851066\\
64	4.35109070474412\\
65	4.35118801383164\\
66	4.35127873941996\\
67	4.35136325710995\\
68	4.35144294702437\\
69	4.35151752314444\\
70	4.35158796973868\\
71	4.35165393336818\\
};
\addlegendentry{Achievable rate (Gauss)};

\end{axis}
\end{tikzpicture}%

%% file: figures_scaled/dis_vs_cont_nondir_sc.tex
%

\begin{tikzpicture}[every node/.style={scale=0.80}]

\pgfplotsset{width=0.4\columnwidth,height =0.45\columnwidth, compat=1.6,
legend image code/.code={
	\draw[mark repeat=1,mark phase=1]
	plot coordinates {
		(0cm,0cm)
		(0cm,0cm)        
		(0.3cm,0cm)         
	};%
}
}
\begin{axis}[%
scale only axis,
separate axis lines,
every outer x axis line/.append style={black},
every x tick label/.append style={font=\color{black}},
xmin=0,
xmax=20,
xlabel={Iteration},
every outer y axis line/.append style={black},
every y tick label/.append style={font=\color{black}},
ymin=0,
ymax=4.5,
ylabel={MI [bpcu]},
axis background/.style={fill=white},
legend style={at={(1,0)},anchor=south east,legend cell align=left,align=left,draw=black,inner sep=1pt,row sep=-2pt,font=\footnotesize}
]
\addplot [color=blue,solid,line width=1.0pt]
  table[row sep=crcr]{%
1	0.0332240904913683\\
2	2.20886329427005\\
3	3.30633365018273\\
4	3.41929724430214\\
5	3.4487189240278\\
6	3.45455627987058\\
7	3.4633973645969\\
8	3.46378954367598\\
9	3.46651176300832\\
10	3.46787453605922\\
11	3.46722788021659\\
12	3.46951255239921\\
13	3.46928454613378\\
14	3.46463414622926\\
15	3.46867088566247\\
16	3.46918536903802\\
17	3.46888979307838\\
18	3.46753630320335\\
19	3.46651100798365\\
20	3.46910438049192\\
21	3.46622412846115\\
22	3.46736068198421\\
23	3.46926507537258\\
24	3.46846666576853\\
25	3.46881175173124\\
26	3.46694164014014\\
27	3.46906660781673\\
28	3.46673270199629\\
29	3.46535397201565\\
30	3.46834817528836\\
31	3.46905045174142\\
32	3.46917843093141\\
33	3.46526293187317\\
34	3.46581670700809\\
35	3.46919633575177\\
36	3.46727201588144\\
37	3.46635470196659\\
38	3.46709701884098\\
39	3.46637511810076\\
40	3.46764105263584\\
41	3.46963303411783\\
42	3.46705531768852\\
43	3.46777647206405\\
44	3.46826750849361\\
45	3.46843529001795\\
46	3.46683820206769\\
47	3.4678589412695\\
48	3.46922875481188\\
49	3.46705088156169\\
50	3.46666530413378\\
51	3.46804785743333\\
52	3.46915259260562\\
53	3.46917515940075\\
54	3.46986416641898\\
55	3.46827723896172\\
56	3.46890790340879\\
57	3.46473837246784\\
58	3.46684727685768\\
59	3.4678115895155\\
60	3.46900085099948\\
61	3.46995823128268\\
62	3.46760018541456\\
63	3.47129528300726\\
64	3.46851372173357\\
65	3.46858624366796\\
66	3.46598950658652\\
67	3.4674200360484\\
68	3.46493686685602\\
69	3.46332768882312\\
70	3.46456559463394\\
71	3.4673538581039\\
};
\addlegendentry{$M=4$, CR-based opt.};

\addplot [color=blue,dotted,line width=1.0pt]
  table[row sep=crcr]{%
1	0.0328588037956033\\
2	2.43822112251022\\
3	3.75636389771314\\
4	3.82193737397603\\
5	3.81783688601493\\
6	3.81148533898348\\
7	3.80498953945804\\
8	3.7973453374288\\
9	3.7909725093735\\
10	3.78726353784903\\
11	3.7834252792356\\
12	3.78218367779407\\
13	3.77870953503819\\
14	3.77845575023783\\
15	3.77864289277602\\
16	3.77683590856997\\
17	3.77659671540736\\
18	3.77627452276919\\
19	3.77546845203924\\
20	3.77571338394315\\
21	3.77506375321852\\
22	3.77435455067796\\
23	3.77415828508504\\
24	3.77423666011004\\
25	3.77433495350624\\
26	3.77517039380129\\
27	3.77524036029496\\
28	3.77514421348632\\
29	3.77325732761135\\
30	3.77294412485388\\
31	3.77406998946485\\
32	3.77349622735037\\
33	3.77556430506881\\
34	3.77541313592746\\
35	3.77512558651458\\
36	3.77395995979346\\
37	3.77630833686389\\
38	3.7743167900054\\
39	3.77248980009434\\
40	3.7729870714703\\
41	3.7746902130888\\
42	3.77244597798405\\
43	3.77318128613085\\
44	3.77413090168727\\
45	3.77505204711054\\
46	3.77184407277237\\
47	3.77342120064073\\
48	3.77318004285648\\
49	3.77258728992727\\
50	3.7723227554811\\
51	3.77168130803163\\
52	3.77175204261183\\
53	3.77362630029883\\
54	3.77172661968006\\
55	3.77174587375317\\
56	3.77084557211232\\
57	3.77346754144586\\
58	3.77279523592206\\
59	3.77342713461627\\
60	3.77207420850602\\
61	3.77379011311205\\
62	3.77136194985518\\
63	3.77300477476312\\
64	3.77245985905999\\
65	3.77187494696657\\
66	3.77151286813459\\
67	3.77182057204304\\
68	3.77314030615393\\
69	3.77316823163527\\
70	3.77203247089148\\
71	3.77296833190453\\
};
\addlegendentry{$M=16$, CR-based opt.};

\addplot [color=red,solid,line width=1.0pt]
  table[row sep=crcr]{%
1	0.0184084325720375\\
2	1.0857139473408\\
3	1.60887464445851\\
4	1.91817155279591\\
5	1.96790807210087\\
6	1.96013208287934\\
7	1.96500500239625\\
8	1.97438434188172\\
9	1.98431608606448\\
10	1.99331630583972\\
11	1.99761191480272\\
12	1.99853054192535\\
13	1.9988654527429\\
14	1.99895584968808\\
15	1.99902030224567\\
16	1.99905954108749\\
17	1.99916819075366\\
18	1.99907836633994\\
19	1.99901142645629\\
20	1.99911289357402\\
21	1.99909572671446\\
22	1.99916472812418\\
23	1.9991080602225\\
24	1.99915885133172\\
25	1.99911582735237\\
26	1.99913494770333\\
27	1.99912448861178\\
28	1.99911596154977\\
29	1.99918326688324\\
30	1.99915354720329\\
31	1.99908897163005\\
32	1.99919280756627\\
33	1.9991840117511\\
34	1.99916444852282\\
35	1.99917820091666\\
36	1.99918940215546\\
37	1.99911589664634\\
38	1.99912405938403\\
39	1.99913024221034\\
40	1.99914635955457\\
41	1.99915815363022\\
};
\addlegendentry{$M=4$, Gauss-based opt.};

\addplot [color=red,dotted,line width=1.0pt]
  table[row sep=crcr]{%
1	0.0205744801156526\\
2	1.12903198109954\\
3	1.67425823692028\\
4	2.15690330813298\\
5	2.57834445922075\\
6	2.91155557637448\\
7	3.13993498531907\\
8	3.2709502377586\\
9	3.34879779445602\\
10	3.39336472665783\\
11	3.42233358873399\\
12	3.44178198643711\\
13	3.4532335036057\\
14	3.46209252124828\\
15	3.46766614318865\\
16	3.47120517850612\\
17	3.47320914318719\\
18	3.47537644795206\\
19	3.47688444860269\\
20	3.47721151471294\\
21	3.47853915552607\\
22	3.47822725433018\\
23	3.47891606181138\\
24	3.47807889251126\\
25	3.47901994773318\\
26	3.47945165676671\\
27	3.47907523481413\\
28	3.4794669565404\\
29	3.47797485979349\\
30	3.47865729674364\\
31	3.47889913814808\\
32	3.4788483844915\\
33	3.47903902648181\\
34	3.47873322559475\\
35	3.47862252951119\\
36	3.47950049446508\\
37	3.47981871188994\\
38	3.47849465154042\\
39	3.480064293033\\
40	3.47990580597825\\
41	3.4796248187689\\
};
\addlegendentry{$M=16$, Gauss-based opt.};

\addplot [color=green,solid,line width=1.0pt]
  table[row sep=crcr]{%
1	0.0198123952761717\\
2	1.14226719305566\\
3	1.8044473616893\\
4	2.39418634898102\\
5	2.86841921383721\\
6	3.19852261687435\\
7	3.43139466677342\\
8	3.5911494013811\\
9	3.69784945291894\\
10	3.76691513496908\\
11	3.8112290432416\\
12	3.84005494532194\\
13	3.85908835030353\\
14	3.87179721635007\\
15	3.88034269199152\\
16	3.88611725942699\\
17	3.8900324377276\\
18	3.89269292168942\\
19	3.89450359067734\\
20	3.89573721602809\\
21	3.89657833497203\\
22	3.89715214761179\\
23	3.8975437629709\\
24	3.89781111562305\\
25	3.8979936800489\\
26	3.89811837116041\\
27	3.89820354956381\\
28	3.89826174505325\\
29	3.89830151078801\\
30	3.89832868672472\\
31	3.89834726105431\\
32	3.89835995781918\\
33	3.89836863788688\\
34	3.89837457263744\\
35	3.89837863081135\\
36	3.89838140609486\\
37	3.89838330425304\\
38	3.89838460264466\\
39	3.89838549087833\\
40	3.89838609858921\\
41	3.89838651441849\\
42	3.89838679898349\\
43	3.89838699374195\\
44	3.89838712705102\\
45	3.89838721830916\\
46	3.89838728078791\\
47	3.89838732356801\\
48	3.89838735286347\\
49	3.89838737292701\\
50	3.89838738666946\\
51	3.89838739608338\\
52	3.89838740253287\\
53	3.89838740695193\\
54	3.89838740998013\\
55	3.89838741205547\\
56	3.89838741347794\\
57	3.89838741445304\\
58	3.89838741512155\\
59	3.8983874155799\\
60	3.89838741589422\\
61	3.89838741610978\\
62	3.89838741625763\\
63	3.89838741635906\\
64	3.89838741642864\\
65	3.89838741647638\\
66	3.89838741650914\\
67	3.89838741653163\\
68	3.89838741654706\\
69	3.89838741655765\\
70	3.89838741656493\\
71	3.89838741656992\\
};
\addlegendentry{Achievable rate (Gauss)};

\end{axis}
\end{tikzpicture}%